\documentclass[preprint,showpacs,preprintnumbers,amsmath,amssymb,showkeys]{revtex4}
 
\usepackage{graphicx}
\usepackage{dcolumn}
\usepackage{bm}
\usepackage{braket}
\usepackage[shortlabels]{enumitem}
  
\begin{document} 

\noindent

\preprint{}

\title{Nonlinear Schr\"odinger equations and generalized Heisenberg uncertainty principle violating the principle of estimation independence}
 
\author{Agung Budiyono$^{a,b,c,d}$} 
\email{agungbymlati@gmail.com}
\author{Hermawan K. Dipojono$^{a,c}$}
\affiliation{$^a$Research Center for Nanoscience and Nanotechnology, Bandung Institute of Technology, Bandung, 40132, Indonesia;}
\affiliation{$^b$Edelstein Center, Hebrew University of Jerusalem, Jerusalem, 91904 Israel;}
\affiliation{$^c$Department of Engineering Physics, Bandung Institute of Technology, Bandung, 40132 Indonesia;}
\affiliation{$^d$Kubus Computing and Research, Juwana, Pati, 59185 Indonesia}
 
\date{\today}
  
\begin{abstract}   

One of the advantages of a reconstruction of quantum mechanics based on transparent physical axioms is that it may offer insight to naturally generalize quantum mechanics by relaxing the axioms. Here, we discuss possible extensions of quantum mechanics within a general epistemic framework based on an operational scheme of estimation of momentum given positions under epistemic restriction. The epistemic restriction is parameterized by a global-nonseparable random variable on the order of Planck constant, an ontic extension to the separable classical phase space variables. Within the estimation scheme, the canonical quantum laws is reconstructed for a specific estimator and estimation error. In the present work, keeping the Born's quadratic law intact, we construct a class of nonlinear variants of Schr\"odinger equation and generalized Heisenberg uncertainty principle within the estimation scheme by assuming a more general class of estimation errors. The nonlinearity of the Schr\"odinger equation and the deviation from the Heisenberg uncertainty principle thus have a common transparent operational origin in terms of generalizations of estimation errors. We then argue that a broad class of nonlinearities and deviations from Heisenberg uncertainty principle arise from estimation errors violating a plausible inferential-causality principle of estimation independence which is respected by the standard quantum mechanics. This result therefore constrains possible extensions of quantum mechanics, and suggests directions to generalize quantum mechanics which comply with the principle of estimation independence.    
  
\end{abstract}    

\maketitle                                       

\section{Introduction}

The linearity of the Schr\"odinger equation with Hermitian quantum Hamiltonian, together with the Born's quadratic law, i.e., the Born's statistical interpretation of wave function, and the Heisenberg uncertainty principle, are the central tenets of quantum mechanics. Hitherto, they have passed all experimental tests with unprecedented accuracy. In spite of their monumental empirical successes, there are nonetheless important reasons to mull over possible generalizations of, and deviations from, these canonical quantum laws: as a guide to conceive stringent precision tests of quantum mechanics which is motivated by the argument that the linearity of the theory might be an approximation to a deeper theory with an extremely weak nonlinearity \cite{Bialynicki-Birula nonlinearity,Weinberg nonlinearity}, or that the Born's quadratic law applies only in a specific situation of quantum equilibrium \cite{Valentini nonequilibrium}; to construct a general framework for a broad class of nonclassical theories which provide a foil to the standard quantum mechanics to better understand conceptually what deeply distinguishes quantum mechanics from the alternative nonclassical theories \cite{Hardy axioms,D'Ariano generalized probabilitstic theory,Dakic-Brukner axioms,Masanes axioms,Paterek axioms,Chiribella axioms,Chiribella-Spekkens quantum axioms proceedings}; to study their information processing capabilities in comparison with those based on quantum mechanics \cite{Barrett axioms,Barnum generalized no-broadcasting,Abrams-Lloyd nonlinearity - fast computation,Aaronson nonlinearity-nonunitary - fast computation,Ver Steeg relaxing uncertainty relation,Barrett computational landscape general physical theories}; to resolve the infamous measurement problem that the present linear Schr\"odinger equation may lead to an embarrassing superposition of perceptible macroscopic objects \cite{De Broglie nonlinearity,Pearle nonlinearity,Gisin nonlinearity-stochasticity,GRW theory,Diosi gravity induced collapse,Bassi collapse model review} (i.e., the well-known Schr\"odinger's cat \cite{Schroedinger's cat}); and to develop a general framework which may encompass quantum mechanics and general relativity \cite{Mielnik generalized quantum mechanics,Hardy - quantum gravity}.  

It has been argued, however, that nonlinear modifications of the Schr\"odinger equation may violate the relativistic causality principle of no-signalling \cite{Gisin nonlinearity - signaling,Polchinski nonlinearity - signaling,Czachor nonlinearity - signaling,Mielnik nonlinearity - signaling,Simon no-signaling imply linearity}. Moreover, introducing a non-Hermitian quantum Hamiltonian \cite{Bender nonHermiticity} may also be in conflict with no-signalling \cite{Lee nonHermitian - signaling}. In contrast to this, while quantum mechanics allows stronger than classical correlation \cite{Bell's theorem,CHSH inequality}, such nonclassical correlations cannot be used to perform faster than light communication; hence, quantum mechanics elegantly respects no-signaling. However, as Popescu-Rohrlich box shows \cite{Popescu-Rohrlich axioms}, quantum mechanics is not the only theory which allows stronger than classical correlation and at the same time also complies with no-signaling. Hence, no-signaling is not sufficient to uniquely single out quantum mechanics from among all possible nonclassical theories. These simple but fundamental results yet suggest that the abstract quantum laws may be deeply rooted in some forms of causality principles. This belief is further supported by the theoretical findings that introducing nonlinearity in the Schr\"odinger equation and a deviation from Heisenberg uncertainty principle may also lead to violations of the second law of thermodynamics \cite{Peres nonlinearity violates 2nd law,Hanggi a deviation from UR violates 2nd} (see however Ref. \cite{Weinberg on nonlinearity and second law}). In addition, a deviation from the Heisenberg uncertainty principle may imply stronger than quantum correlation \cite{Oppenheim-Wehner entropic UR and QS}, which in turn allows implausible computational power \cite{Popescu review,Dam informational approach Tsirelson bound,Brassard informational approach Tsirelson bound,Buhrman superstrong cryptography,Linden nonlocal computation,Brunner trivial communication,Pawlowski informational approach Tsirelson bound,Gross trivial dynamics with superstrong correlation}. Is quantum mechanics the unique nonclassical theory that obeys certain causality principles \cite{Popescu-Rohrlich axioms}? This line of inquiry to reconstruct quantum mechanics from deep but simple and transparent axioms \cite{Wheeler: howcome quantum}, may not only lead to a better understanding on the meaning of quantum mechanics, it may also offer fresh insight and useful intuition to suggest a logically coherent possible extensions of quantum mechanics by slightly varying the parameters unfixed by the axioms or by relaxing some of the axioms.     

On the other hand, previously, we have shown that the abstract formalism of nonrelativistic spinless quantum mechanics can be reconstructed within an epistemic framework based on an operational scheme of estimation of momentum given the information on the conjugate positions \cite{Agung epistemic interpretation}, under a fundamental epistemic restriction \cite{Spekkens toy model with epistemic restriction} so that the allowed probability distribution of positions that an agent can prepare are irreducibly parametrized by the underlying momentum field. The momentum field is assumed to fluctuate randomly induced by a global-nonseparable random variable on the order of Planck constant, an ontic extension to the separable classical phase space variables \cite{Agung-Daniel model}. Within this operational scheme of estimation under epistemic restriction, we showed in Refs. \cite{Agung epistemic interpretation,Agung-Daniel model} that the mathematical rules of quantum mechanics in complex Hilbert space formalism, including the linear Schr\"odinger equation with Hermitian quantum Hamiltonian and Born's quadratic law, and also the exact form of the Heisenberg uncertainty principle, emerge when the associated estimator and estimation error, take `specific' forms. Such a specific operational scheme of estimation of momentum given positions has a well-defined implementation in terms of weak momentum value measurement \cite{Aharonov weak value,Lundeen complex weak value,Jozsa complex weak value}, which leads to a simple method for the reconstruction of quantum wave function \cite{Agung epistemic interpretation,Agung ERPS distribution}.  

In the present work, we show that the above epistemic framework based on the operational scheme of estimation of momentum given positions, is flexible to transparently accommodate a broad class of possible extensions of quantum mechanics. Keeping the Born's quadratic law intact, we first construct a broad class of nonlinear variants of the Schr\"odinger equation and generalized Heisenberg uncertainty principle, by choosing a general class of estimation errors. Both deviations from the canonical laws of quantum mechanics have thus a common origin from, and a transparent operational meaning in terms of, the generalizations of the estimation errors. They are thus deeply interrelated. In particular, there is no nonlinearity without a deviation from the Heisenberg uncertainty principle, suggesting that it is difficult to modify a part of quantum mechanics without changing the other important parts of the theory. We then show that a broad class of nonlinear Schr\"odinger equations and deviations from Heisenberg uncertainty principle arise from estimation errors violating a plausible inferential-causality principle of estimation independence \cite{Agung estimation independence}. By contrast, the principle of estimation independence is strictly and pleasingly respected by the specific estimation error leading to the standard quantum mechanics. The result thus constrains possible extensions of quantum mechanics, and offers insight to the kinds of generalization of quantum mechanics which comply with the principle of estimation independence.  

The rest of the paper is organized as follows. In Sec. \ref{An epistemic reconstruction of quantum mechanics: estimation under epistemic restriction} we give a brief summary on the epistemic reconstruction of nonrelativistic spinless quantum mechanics proposed in Refs. \cite{Agung-Daniel model,Agung epistemic interpretation,Agung estimation independence}, based on the operational scheme of estimation of momentum given positions under epistemic restriction parameterized by a global random variable, with `specific' estimator and estimation error. In Sec. \ref{Modified estimation errors} we consider a generalization of the estimation scheme by employing a class of more general estimation errors, based on which we derive a broad nonlinear variants of Schr\"odinger equation in Sec. \ref{A class of nonlinear Schroedinger equation}, and generalized Heisenberg uncertainty principle in Sec. \ref{A class of modified uncertainty relations}. We proceed in Sec. \ref{discussion} to discuss the relation between the resulting nonlinearity in the Schr\"odinger equation and the deviation from the Heisenberg uncertainty principle, and introduce a physically transparent and plausible inferential-causality principle of estimation independence \cite{Agung estimation independence} to rule out a large class of nonlinearities and deviations from the Heisenberg uncertainty principle. We end in Sec. \ref{conclusion} with conclusions  and offers a sketch on various future directions for possible generalizations of quantum mechanics which do not violate the principle of estimation independence. 

\section{Quantum mechanics from a specific scheme of estimation under epistemic restriction parameterized by a global random variable on the order of Planck constant\label{An epistemic reconstruction of quantum mechanics: estimation under epistemic restriction}}

Consider a system with a spatial configuration $q=(q_1,\dots,q_N)$ and the conjugate momentum $p=(p_1,\dots,p_N)$. First, recall that in classical mechanics, working within the Hamilton-Jacobi formalism \cite{Rund book: Hamilton-Jacobi formalism}, the momentum field can be written as 
\begin{eqnarray}
\tilde{p}_{\rm C}(q,t)=\partial_qS_{\rm C}(q,t), 
\label{HJ condition}
\end{eqnarray}
where $\partial_q=(\partial_{q_1},\dots,\partial_{q_N})$, and $S_{\rm C}(q,t)$ is a real-valued scalar function of the positions $q$ and time $t$, called as the Hamilton's principal function. (In the paper, we label the momentum field with $\tilde{p}$, whereas $p$ is used to denote the specific value of momentum.) It is then clear from Eq. (\ref{HJ condition}) that in classical mechanics, given a momentum field $\tilde{p}_{\rm C}(q)$  (trivial dependence on time is notationally supressed) arising in a fixed experimental arrangement, it is in principle possible for an agent, by repeating the experiments many times, to prepare an ensemble of trajectories with arbitrary distribution of positions $\rho(q)$. Namely, each trajectory in the momentum field $\tilde{p}_{\rm C}(q)$ can be assigned an arbitrary weight $\rho(q)$. Hence, in classical mechanics, the distribution of positions $\rho(q)$ is fundamentally independent of, thus is not irreducibly parametrized by, the underlying momentum field $\tilde{p}_{\rm C}(q)$. 

We postulate that the above `epistemic freedom', namely the freedom to prepare the probability distribution of positions independent of the underlying momentum field, is no longer respected in microscopic world \cite{Agung-Daniel model}. Assume first that in microscopic world, there is a global-nonseparable variable $\xi$ of action dimensional, fluctuating randomly inducing a random fluctuations of the momentum field $\tilde{p}(q,t;\xi)=\big(\tilde{p}_1(q,t;\xi),\dots,\tilde{p}_N(q,t;\xi)\big)$. We then assume that the ensemble of trajectories obtained by identically repeating the experiment suffers a fundamental `epistemic restriction' \cite{Agung-Daniel model}: namely, unlike in classical mechanics discussed above, it is no longer possible for an agent to assign each trajectory in the momentum field $\tilde{p}(q;\xi)$ an arbitrary weight. The probability distributions of positions therefore fundamentally depends on, thus irreducibly parametrized by, the underlying momentum field $\tilde{p}(q;\xi)$. To make explicit this intrinsic dependence, we write the probability distribution of positions as $\rho_{\tilde{p}}(q)$ with a subscript $\tilde{p}$. Furthermore, we assume that in the formal limit of vanishing global fluctuation $\xi$, the epistemic restriction disappears, i.e., $\lim_{\xi\rightarrow 0}\rho_{\tilde{p}}(q)=\rho(q)$, and we regain classical mechanics satisfying Eq. (\ref{HJ condition}) with the epistemic freedom recovered. The fluctuation of $\xi$ thus characterizes the strength of the epistemic restriction, and therefore must be practically ignorable in the macroscopic physical regime. We emphasize that the global-nonseparable variable $\xi$ provides an ontic extension to the separable classical phase space variables. 

We have argued in Refs. \cite{Agung-Daniel model,Agung epistemic interpretation} that the abstract mathematical rules of nonrelativistic spinless quantum mechanics can be derived within an operational scheme of estimation of momentum given positions under the above epistemic restriction, combined with the Bayesian reasoning given the experimental settings. A concrete illustration of the reconstruction of quantum laws based on such an estimation scheme in a single and double slits experiment is given in the Appendix \ref{A concrete illustration of the epistemic reconstruction of quantum mechanics within the operational scheme of estimation under epistemic restriction}. First, suppose that the agent has access to $q$ via some position measurement. Note that as emphasized by Bell \cite{Bell speakable book}, any measurement should be reducible to the measurement of position. Since $q$ is sampled from $\rho_{\tilde{p}}(q)$ parametrized by $\tilde{p}(q;\xi)$, then it must somehow contain some information about $\tilde{p}(q;\xi)$. How can the agent use her information about position, in the most reasonable way, to estimate the conjugate momentum? To answer this parameter estimation problem, we need to choose the estimator and the associated estimation error \cite{Papoulis and Pillai book on probability and statistics}. 

Let us construct a reasonable estimator for $\tilde{p}(q;\xi)$. First, we select a sub-ensemble of trajectories that are passing $q(t)$ at time $t$, where different trajectories in the sub-ensemble correspond to different fluctuations of $\xi$. Then, along each of the trajectory in the sub-ensemble, we make a `naive classical' momentum measurement via two consecutive position measurements as follows. Just before the system is detected at $q(t)$, we perform a sufficiently weak measurement of the position at time $t-\Delta t$ without appreciably disturbing the trajectory, yielding $q(t-\Delta t)$, where $\Delta t$ is extremely small. The velocity along the trajectory at $q(t)$ can then be computed in the conventional way by evaluating the difference between $q(t)$ and $q(t-\Delta t)$ and dividing it with $\Delta t$, from which we also get the momentum $\tilde{p}(q;\xi)$ along that particular trajectory. Note that, because of the fluctuation of $\xi$, each such single measurement of momentum must yield a random outcome. We then define the estimator $\overline{p}(q)=(\overline{p}_1(q),\dots,\overline{p}_N(q))$ for $\tilde{p}(q;\xi)$ at time $t$ by taking the average of the above measurement outcomes over all the trajectories in the sub-ensemble. Within the statistical model, such a conditional ensemble average of momentum $\overline{p}(q)$ thus corresponds to the average of $\tilde{p}(q;\xi)$ over $\xi$, i.e., 
\begin{eqnarray}
\overline{p}(q)\doteq\int{\rm d}\xi~\tilde{p}(q;\xi)\chi(\xi), 
\label{estimate of momentum based on mean average}
\end{eqnarray}
where $\chi(\xi)$ is the probability distribution of $\xi$. Clearly, by construction, in the absence of $\xi$, the above scheme for estimating the momentum reduces to the conventional measurement of momentum at $q$ in classical mechanics which must give back Eq. (\ref{HJ condition}). 

Next, to have a smooth correspondence with classical mechanics, we assume that the above estimator $\overline{p}(q)$ for $\tilde{p}(q;\xi)$ at $q$ can be written as follows: 
\begin{eqnarray}
\overline{p}(q)\doteq\partial_qS(q), 
\label{weakly unbiased best estimator}
\end{eqnarray}
where $S(q)$ is a real-valued scalar function, so that in the macroscopic physical regime, the estimator is expected to approach the gradient of the Hamilton's principal function, i.e., $\overline{p}(q)=\partial_qS (q)\rightarrow \partial_qS_{\rm C}(q)$, recovering Eq. (\ref{HJ condition}) of classical mechanics. Of course, since we want to reconstruct quantum mechanics from the above estimation scheme, for consistency, we need to check afterward whether the above operational protocol for estimating the momentum at $q(t)$, by first weakly measuring the position at time $t-\Delta t$ and then followed immediately by a position post-selection (strong position measurement) at time $t$, is consistent with quantum mechanics. That this is indeed the case is shown by Wiseman in Ref. \cite{Wiseman Bohmian velocity from naive weak value measurement} (see also Refs. \cite{Agung ERPS distribution,Agung epistemic interpretation}), which has led to the impressive experimental reconstruction of the average trajectory in the double slits experiment \cite{Steinberg average trajectory}. Namely, implementing the above estimation of the momentum at $q$ via an ensemble of two successive position measurements, with the quantum weak measurement over a pre-selected wave function $\psi(q)$ and a position post-selection at $q$ \cite{Aharonov weak value,Lundeen complex weak value,Jozsa complex weak value}, indeed yields Eq. (\ref{weakly unbiased best estimator}), where $S(q)$ is identified as the phase of quantum wave function $\psi(q)$.  

Moreover, given $q$, let us assume that the error in a single-shot estimation of $\tilde{p}(q;\xi)$ with the estimator $\overline{p}(q)=\partial_qS(q)$ has the following `specific' form \cite{Agung epistemic interpretation}:
\begin{eqnarray}
\epsilon_p(q;\xi)\doteq \tilde{p}(q;\xi)-\partial_qS(q)=\frac{\xi}{2}\partial_q\ln\rho_{\tilde{p}}(q).
\label{estimation error}
\end{eqnarray}
One can see that in the mathematical limit $\xi\rightarrow 0$, the estimation error is vanishing, and we regain the classical relation of Eq. (\ref{HJ condition}), $\lim_{\xi\rightarrow 0}\tilde{p}=\overline{p}=\partial_qS$, so that the epistemic restriction disappears, as required. Furthermore, assuming that $\rho_{\tilde{p}}(q)$ is vanishing at the boundary, the above estimation error is on average vanishing for all $\xi$, i.e., $\int{\rm d}q\epsilon_p(q;\xi)\rho_{\tilde{p}}(q)=\frac{\xi}{2}\int{\rm d}q\partial_q\rho_{\tilde{p}}(q)=0$, ${\rm d}q={\rm d}q_1\dots{\rm d}q_N$; hence, it is desirably (weakly) unbiased. 

Let us further assume that the global variable $\xi$ is fluctuating randomly on a microscopic timescale so that its first and second moments are independent of time, given by \cite{Agung-Daniel model} 
\begin{equation}
\overline{\xi}\doteq\int {\rm d}\xi~\xi~\chi(\xi)=0,~~\overline{\xi^2}=\hbar^2.
\label{Planck constant}
\end{equation} 
The left equation guarantees that the conditional (sub-ensemble) average of $p$ given $q$ is equal to the estimator satisfying Eq. (\ref{weakly unbiased best estimator}); i.e., from Eq. (\ref{estimation error}), we have: $\overline{p}(q)=\int{\rm d}\xi\tilde{p}(q;\xi)\chi(\xi)=\partial_qS(q)$. On the other hand, the right equation in Eq. (\ref{Planck constant}) shows that the strength of the estimation error is on the order of Planck constant. It therefore ensures that in the macroscopic physical regime, the estimation error is much smaller than the estimator, i.e., $|\partial_qS|\gg|\frac{\xi}{2}\partial_q\ln\rho_{\tilde{p}}|$, so that Eq. (\ref{estimation error}) effectively reduces back to the classical relation: $\tilde{p}\approx \partial_qS$. Finally, one can also argue that in the above estimation scheme, the estimator $\overline{p}(q)=\partial_qS(q)$ ``best'' estimates $\tilde{p}(q;\xi)$, in the sense that it minimizes the mean-squared (MS) error defined as $\mathcal{E}_p^2\doteq\int{\rm d}q{\rm d}\xi\big(\epsilon_p(q;\xi)\big)^2\chi(\xi)\rho_{\tilde{p}}(q)$ \cite{Agung epistemic interpretation} (see also Appendix \ref{proof of best estimator}). This estimation scheme is thus also consistent with the argument advanced in Refs. \cite{Hall weak value as optimal estimate,Johansen weak value best estimation} wherein Eq. (\ref{weakly unbiased best estimator}), with $S(q)$ is given by the phase of the wave function, is interpreted as the optimal estimate of momentum based on the measurement of position. 

Next, for later comparison, let us write Eq. (\ref{estimation error}) as 
\begin{eqnarray}
\tilde{p}(q;\xi)=\partial_qS(q)+\frac{\xi}{2}\frac{\partial_q\rho_{\tilde{p}}(q)}{\rho_{\tilde{p}}(q)}. 
\label{fundamental epistemic decomposition}
\end{eqnarray}
Hence, we have a random momentum field which is decomposed into two terms. We emphasize that, by construction, the above decomposition of the random momentum field is not ontic (physical) happening in physical space. Rather, the decomposition is epistemic (i.e., informational); namely, it happens in the agent's mind, artificially devised by the agent to describe her best estimate of the momentum given positions (the first term on the right-hand side of Eq. (\ref{fundamental epistemic decomposition})) and the associated single-shot estimation error (the second term) \cite{Agung epistemic interpretation}. Equation (\ref{fundamental epistemic decomposition}) is just the specific epistemic restriction we postulated in Ref. \cite{Agung-Daniel model}, based on which we derived the mathematical formalism of quantum mechanics. 

Within the epistemic reconstruction based on the specific operational scheme of estimation of momentum given positions, the quantum wave function $\psi(q,t)$ characterizing a preparation is a mathematical object which summarizes the estimator of Eq. (\ref{weakly unbiased best estimator}) and the estimation error of Eq. (\ref{estimation error}) via $\big(S(q,t),\rho_{\tilde{p}}(q,t)\big)$ as \cite{Agung epistemic interpretation}
\begin{equation} 
\psi(q,t)\doteq\sqrt{\rho_{\tilde{p}}(q,t)}\exp(iS(q,t)/\hbar). 
\label{wave function}
\end{equation}
As in Refs. \cite{Wiseman Bohmian velocity from naive weak value measurement,Hall weak value as optimal estimate,Johansen weak value best estimation}, $S(q,t)$ operationally defined in Eq. (\ref{weakly unbiased best estimator}) indeed constitutes the phase of the quantum wave function. In this sense, basically, the estimation of momentum given position described above thus operationally leads to the reconstruction of quantum wave function characterizing the preparation \cite{Agung ERPS distribution}. For example, consider a preparation setting so that quantum mechanically it results in a Gaussian wave function $\psi(q)=(\frac{1}{2\pi\sigma_q^2})^{1/4} e^{-(q-q_o)^2/4\sigma_q^2+ip_oq/\hbar}$. Within the above epistemic interpretation, noting Eqs. (\ref{weakly unbiased best estimator}) and (\ref{estimation error}), it means that given information on $q$, the agent should assign $\overline{p}(q)=p_o$ as her best estimate of the momentum of the system, with the single-shot estimation error $\epsilon_p(q;\xi)=-\frac{\xi}{2\sigma_q^2}(q-q_o)$ so that the MS error reads $\mathcal{E}_p^2=\hbar^2/4\sigma_q^2$. In particular, a preparation leading to a plane wave function, $\psi(q)\sim e^{ip_oq/\hbar}$, means that the agent's best estimate of momentum $\overline{p}=p_o$ is sharp with a vanishing MS error, $\mathcal{E}_p^2=0$. 

Hence, by construction, quantum wave function is not an agent-independent objective physical attribute of the system, but it represents the agent's estimation about the momentum field arising in her preparation based on information on the conjugate positions \cite{Agung epistemic interpretation}. Note that from the definition of wave function in Eq. (\ref{wave function}), the epistemic decomposition of momentum field in Eq. (\ref{fundamental epistemic decomposition}) is invariant under the transformation of wave function $\psi\mapsto Z\psi$, where $Z$ is an arbitrary complex constant. Namely, the estimator and the estimation error of Eqs. (\ref{weakly unbiased best estimator}) and (\ref{estimation error}) are invariant under such transformation of wave function. $\psi$ and $Z\psi$ thus represent the same estimation scheme, i.e., the statistical content encoded in $\psi$ and $Z\psi$ are the same, as in standard quantum mechanics. One can also see that, by construction, Eq. (\ref{wave function}) leads to the Born's quadratic law 
\begin{equation} 
\rho_{\tilde{p}}(q,t)=|\psi(q,t)|^2. 
\label{Born's quadratic law}
\end{equation} 

Finally, within the above specific estimation scheme, the linear Schr\"odinger equation can be seen as a Bayesian rule for updating the specific estimator and estimation error represented by the wave function via Eqs. (\ref{weakly unbiased best estimator}) and (\ref{estimation error}), when she does not make measurement  \cite{Agung epistemic interpretation}. To see this, first, note that measurement is in practice carried out by making a selection of a sub-ensemble of trajectories associated with a particular measurement outcome (see Appendix \ref{A concrete illustration of the epistemic reconstruction of quantum mechanics within the operational scheme of estimation under epistemic restriction} for a concrete illustration). No measurement thus corresponds to no selection of trajectories. In the absence of measurement, it is therefore natural for the agent to update her estimation represented by the wave function by imposing the statistical-informational constraints of conservation of trajectories and average energy. It is shown in Ref. \cite{Agung-Daniel model} that, within the estimation scheme with the specific estimator and estimation error given by Eqs. (\ref{weakly unbiased best estimator}) and (\ref{estimation error}), the above conservation principles lead to the derivation of the celebrated linear Schr\"odinger equation. We shall rederive the linear Schr\"odinger equation as a specific case of a more general dynamical equation in Sec. \ref{A class of nonlinear Schroedinger equation}. Moreover, the Heisenberg-Kennard uncertainty relation between momentum and position can be traced back to the trade-off between the MS errors of simultaneous estimations of momentum field and mean position, which in turn is implied by the specific choice of estimation error of Eq. (\ref{estimation error}). This fundamentally distinctive feature of quantum mechanics will also be rederived in Sec. \ref{A class of modified uncertainty relations} as a specific case of a more general uncertainty relation. 

\section{Generalized estimation errors: nonlinear Schr\"odinger equation, and generalized Heisenberg uncertainty principle \label{Nonlinear Schroedinger equation and modified Heisenberg uncertainty principle within an estimation scheme}} 

\subsection{A class of generalized estimation errors\label{Modified estimation errors}}

One of the advantages of the epistemic reconstruction of quantum mechanics within the operational scheme of estimation under epistemic restriction is that, it provides a flexible operational framework for transparently accommodating a broad class of possible generalizations of quantum mechanics. As summarized above, since the exact forms of the linear Schr\"odinger equation and the Heisenberg uncertainty principle can be obtained starting from the scheme of estimation of momentum given positions with the help of `specific' estimator and estimation error respectively given by Eqs. (\ref{weakly unbiased best estimator}) and (\ref{estimation error}), it is instructive to generalize the above estimation scheme by relaxing Eqs. (\ref{weakly unbiased best estimator}) or/and (\ref{estimation error}), to search for possible nontrivial extensions of quantum mechanics. To this end, recall that, as discussed in the previous section, the choice of the estimator of Eq. (\ref{weakly unbiased best estimator}) is primarily motivated by a desire to have a smooth macroscopic classicality, requiring the estimator to recover the classical relation of Eq. (\ref{HJ condition}) in the macroscopic physical regime. In this sense, the form of the estimator of Eq. (\ref{weakly unbiased best estimator}) appears to be very natural. By contrast, the form of the estimation error of Eq. (\ref{estimation error}) appears to be apparently ad-hoc. Hence, it is instructive to try various possible alternative forms of estimation error, and work out and analyze the modifications they imply to the canonical laws of standard quantum mechanics such as the linear Schr\"odinger equation and the Heisenberg uncertainty principle.    

Let us therefore consider a generalized scheme of estimation of the momentum based on information on the conjugate positions, with the estimator given by Eq. (\ref{weakly unbiased best estimator}), but with an estimation error which generalizes Eq. (\ref{estimation error}) having the following general form:
\begin{eqnarray}
\epsilon_{p_f}(q;\xi)&\doteq&\tilde{p}(q;\xi)-\partial_qS(q)\nonumber\\
&=&\frac{\xi}{2}\frac{\partial_q\rho_{\tilde{p}}(q)}{\rho_{\tilde{p}}(q)}+\frac{\xi}{2}f\big(\rho_{\tilde{p}}(q),\partial_q\rho_{\tilde{p}}(q)\big), 
\label{generalized estimation error}
\end{eqnarray}
where $\xi$ is again assumed to satisfy Eq. (\ref{Planck constant}), and $f=\big(f_1(\rho_{\tilde{p}},\partial_q\rho_{\tilde{p}}),\dots,f_N(\rho_{\tilde{p}},\partial_q\rho_{\tilde{p}})\big)$ is a real vector-valued function of $\rho_{\tilde{p}}(q)$ and its spatial gradient $\partial_q\rho_{\tilde{p}}(q)$. Generalization to include higher degrees of spatial derivatives of $\rho_{\tilde{p}}(q)$ are straightforward. Comparing Eq. (\ref{generalized estimation error}) with Eq. (\ref{estimation error}), we have thus added a minimal yet general nontrivial correction term given by the last term on the right-hand side of Eq. (\ref{generalized estimation error}). 

Several desirable properties of the specific estimation scheme of Sec. \ref{An epistemic reconstruction of quantum mechanics: estimation under epistemic restriction} are shared by the above more general estimation scheme. First, in the limit of vanishing global fluctuation $\xi$, the estimation error of Eq. (\ref{generalized estimation error}) is vanishing, and we consistently recover Eq. (\ref{HJ condition}) of classical mechanics, i.e., $\lim_{\xi\rightarrow 0}\tilde{p}=\overline{p}=\partial_qS$. Next, in the macroscopic regime where the estimation error of Eq. (\ref{generalized estimation error}) is much smaller than the estimator of Eq. (\ref{weakly unbiased best estimator}), we again effectively regain the classical relation of Eq. (\ref{HJ condition}), i.e., $\tilde{p}\approx\partial_qS(q)$. Moreover, noting Eq. (\ref{Planck constant}), from Eq. (\ref{generalized estimation error}), the conditional average of $p$ given $q$ is equal to the estimator of Eq. (\ref{weakly unbiased best estimator}), i.e., $\int{\rm d}\xi\tilde{p}(q;\xi)\chi(\xi)=\partial_qS(q)$, as required. Finally, as shown in Appendix \ref{proof of best estimator}, like the specific scheme of estimation in Sec. \ref{An epistemic reconstruction of quantum mechanics: estimation under epistemic restriction}, in the estimation scheme with the general estimation error of Eq. (\ref{generalized estimation error}), the estimator of Eq. (\ref{weakly unbiased best estimator}) also provides the best estimate of momentum given positions, minimizing the MS error.  

We show below that the general form of estimation error of Eq. (\ref{generalized estimation error}) will lead to a broad class of nonlinear variants of Schr\"odinger equation when the agent does not make measurement (Sec. \ref{A class of nonlinear Schroedinger equation}), and a class of generalized Heisenberg uncertainty principle (Sec. \ref{A class of modified uncertainty relations}). We note that in Ref. \cite{Agung estimation independence} we have also briefly discussed a specific modification of estimation error of Eq. (\ref{estimation error}) leading to a specific deviation from the Heisenberg uncertainty principle; this specific modification belongs to the class of estimation errors of Eq. (\ref{generalized estimation error}) with a specific $f=\Lambda\partial_q\rho_{\tilde{p}}(q)$, where $\Lambda$ is a dimensionless real constant.  

\subsection{A class of nonlinear Schr\"odinger equations \label{A class of nonlinear Schroedinger equation}} 

Let us derive the equation that governs the time evolution of the agent's estimation of momentum given positions, namely the time evolution of the estimator and estimation error respectively given by Eqs. (\ref{weakly unbiased best estimator}) and (\ref{generalized estimation error}), when the agent does not make any selection of trajectories. We thus need to find out how the agent should rationally update the pair of functions $S(q,t)$ and $\rho_{\tilde{p}}(q,t)$ which determine the estimator and estimation error, provided that she does not make any selection of trajectories. To do this, first, we rewrite Eq. (\ref{generalized estimation error}) as 
\begin{eqnarray}
\tilde{p}(q;\xi)=\partial_qS(q)+\frac{\xi}{2}\frac{\partial_q\rho_{\tilde{p}}(q)}{\rho_{\tilde{p}}(q)}+\frac{\xi}{2}f\big(\rho_{\tilde{p}}(q),\partial_q\rho_{\tilde{p}}(q)\big). 
\label{generalized fundamental epistemic decomposition}
\end{eqnarray}
As for the case of Eq. (\ref{fundamental epistemic decomposition}), by construction, the above decomposition of the random momentum field on the left-hand side, into three terms on the right-hand side, is not ontic happening in physical space; rather, it is epistemic, artificially constructed in the agent's mind to organize her experiences. 

Now, for simplicity, we confine our discussion to a system of $N$ one-dimensional (or $N/3$ three-dimensional) particles subjected to a scalar potential $V(q)$ with the classical Hamiltonian taking the following form: $H(p,q)=\sum_{j=1}^Np_j^2/2m_j+V(q)$, where $m_j$ is the mass of the $j-$th particle. (Application to more general classical Hamiltonian can be done following the same steps below.) In this case, the velocity $\dot{q_j}={\rm d}q_j/{\rm d}t$, $j=1,\dots,N$ and the momentum are related as $\dot{q}_j=\partial H/\partial p_j=p_j/m_j$, $j=1,\dots,N$, so that inserting Eq. (\ref{generalized fundamental epistemic decomposition}), the velocity field is epistemically decomposed as $\tilde{\dot{q}}_j(q;\xi)=\tilde{p}_j/m_j=\frac{\partial_{q_j}S}{m_j}+\frac{\xi}{2m_j}\frac{\partial_{q_j}\rho_{\tilde{p}}}{\rho_{\tilde{p}}}+\frac{\xi}{2m_j}f_j\big(\rho_{\tilde{p}},\partial_q\rho_{\tilde{p}}\big)$, $j=1,\dots,N$. The first term on the right-hand side is just the agent's best estimate of the velocity given positions, and the other two terms comprise the estimation error. Hence, averaging over $\xi$, and noting Eq. (\ref{Planck constant}), the conditional average velocity at $q$ is equal to the best estimate, i.e.,  
\begin{eqnarray}
\overline{\dot{q}_j}(q)=\partial_{q_j}S(q)/m_j,
\label{average velocity - weakly unbiased best estimate of velocity}
\end{eqnarray} 
$j=1,\dots,N$. 

Next, since the agent does not make any selection of trajectories, it is reasonable to require that her estimator and estimation error should be updated in such a way that they respect the conservation of trajectories or probability current. The agent's estimation should therefore satisfy the following continuity equation: $\partial_t\rho_{\tilde{p}}+\sum_{j=1}^N\partial_{q_j}\big(\overline{\dot{q}_j}\rho_{\tilde{p}}\big)=0$. Inserting Eq. (\ref{average velocity - weakly unbiased best estimate of velocity}), one thus obtains
\begin{eqnarray}
\partial_t\rho_{\tilde{p}}+\sum_{j=1}^N\partial_{q_j}\Big(\frac{\partial_{q_j}S}{m_j}\rho_{\tilde{p}}\Big)=0. 
\label{continuity equation}
\end{eqnarray}
Moreover, note that since the underlying momentum field is random due to the fluctuation of $\xi$, each single trajectory does not in general conserve the energy. However, since the agent does not make any selection of trajectories, it is reasonable to assume that her estimation should respect a weaker constraint of  conservation of {\it average} energy, i.e., 
\begin{eqnarray}
\frac{{\rm d}}{{\rm d}t}\braket{H}_{\{S,\rho_{\tilde{p}}\}}=0. 
\label{conservation of average energy}
\end{eqnarray}
Here, the average energy $\braket{H}_{\{S,\rho_{\tilde{p}}\}}$ is defined as in conventional probability theory, i.e., $\braket{H}_{\{S,\rho_{\tilde{p}}\}}=\int{\rm d}q{\rm d}\xi{\rm d}pH(p,q){\rm P}(p,q|\xi)\chi(\xi)$, where ${\rm P}(p,q|\xi)=\prod_{j=1}^N\delta\big(p_j-\tilde{p}_j(q;\xi)\big)\rho_{\tilde{p}}(q)$ is ``the epistemically restricted phase-space distribution'' induced by the momentum field $\tilde{p}(q;\xi)$ defined in Eq. (\ref{generalized fundamental epistemic decomposition}) \cite{Agung-Daniel model,Agung ERPS distribution}.

We show below that the above two reasonable statistical-informational constraints for updating of the agent's estimation of the momentum field when she does not make any selection of trajectories, i.e., the conservation of trajectories and average energy respectively mathematically expressed by Eqs. (\ref{continuity equation}) and (\ref{conservation of average energy}), are sufficient to deduce the time evolution of $S(q,t)$ and $\rho_{\tilde{p}}(q,t)$, which in turn determines the time evolution of the agent's estimator and estimation error via respectively Eqs. (\ref{weakly unbiased best estimator}) and (\ref{generalized estimation error}). First, to solve Eq. (\ref{conservation of average energy}), we must first compute the ensemble average energy, using Eq. (\ref{generalized fundamental epistemic decomposition}), to obtain
\begin{eqnarray}
&&\braket{H}_{\{S,\rho_{\tilde{p}}\}}\nonumber\\
&\doteq&\int{\rm d}q{\rm d}\xi{\rm d}pH(p,q)\prod_{j=1}^N\delta\big(p_j-\tilde{p}_j(q;\xi)\big)\chi(\xi)\rho_{\tilde{p}}(q)\nonumber\\
&=&\sum_{j=1}^N\int{\rm d}q\rho_{\tilde{p}}(q)\Big(\frac{(\partial_{q_j}S)^2}{2m_j}+V+\frac{\hbar^2}{8m_j}\Big(\frac{\partial_{q_j}\rho_{\tilde{p}}}{\rho_{\tilde{p}}}\Big)^2\Big)\nonumber\\
&+& D_f[\rho_{\tilde{p}}],~~~~~~~~~~~~~~~~~~~~~~~~~~~~~~~ 
\label{average energy}
\end{eqnarray}
where we have used Eq. (\ref{Planck constant}), and $ D_f$ is a functional of $\rho_{\tilde{p}}(q)$ defined as 
\begin{eqnarray}
 D_f[\rho_{\tilde{p}}]\doteq\sum_{j=1}^N\int{\rm d}q\Big(\frac{\hbar^2}{4m_j}\frac{\partial_{q_j}\rho_{\tilde{p}}}{\rho_{\tilde{p}}}f_j+\frac{\hbar^2}{8m_j}f_j^2\Big)\rho_{\tilde{p}}(q).
\label{average energy correction}
\end{eqnarray}
Taking the total derivative of Eq. (\ref{average energy}) with respect to time, one gets
\begin{eqnarray}
\frac{{\rm d}}{{\rm d}t}\braket{H}_{\{S,\rho_{\tilde{p}}\}}&=&\sum_{j=1}^N\int{\rm d}q~\partial_t\rho_{\tilde{p}}(q)\Big(\partial_tS+\frac{(\partial_{q_j}S)^2}{2m_j}+V\nonumber\\
&-&\frac{\hbar^2}{2m_j}\frac{\partial_{q_j}^2\sqrt{\rho_{\tilde{p}}}}{\sqrt{\rho_{\tilde{p}}}}+\mathcal{N}_f\big(\rho_{\tilde{p}}\big)\Big), 
\label{modified HJM equation 0}
\end{eqnarray}
where we have made use of Eq. (\ref{continuity equation}), and $\mathcal{N}_f$ is defined as the functional derivative of $ D_f[\rho_{\tilde{p}}]$ with respect to $\rho_{\tilde{p}}(q)$ as
 \begin{eqnarray}
 \mathcal{N}_f\big(\rho_{\tilde{p}}(q)\big)\doteq\frac{\delta  D_f}{\delta \rho_{\tilde{p}}(q)}. 
 \label{correction to HJM and S equations}
 \end{eqnarray}
See Appendix \ref{derivation of HJM equation} for the straightforward derivation. Equating the right-hand side of Eq. (\ref{modified HJM equation 0}) to zero, i.e., imposing the conservation of average energy of Eq. (\ref{conservation of average energy}), one thus obtains the following equation: 
\begin{eqnarray}
\partial_tS+\sum_{j=1}^N\Big(\frac{(\partial_{q_j}S)^2}{2m_j}-\frac{\hbar^2}{2m_j}\frac{\partial_{q_j}^2\sqrt{\rho_{\tilde{p}}}}{\sqrt{\rho_{\tilde{p}}}}\Big)+V\nonumber\\
+\mathcal{N}_f\big(\rho_{\tilde{p}}\big)=0. 
\label{modified HJM equation}
\end{eqnarray}

Hence, to comply with the conservation of trajectories and average energy, the agent's estimation of the momentum given positions with the associated estimator and estimation error determined by $(S(q),\rho_{\tilde{p}}(q))$ via Eqs. (\ref{weakly unbiased best estimator}) and (\ref{generalized estimation error}), must satisfy a pair of differential equations, i.e., Eqs. (\ref{continuity equation}) and (\ref{modified HJM equation}). Finally, defining the wave function as in Eq. (\ref{wave function}), the two coupled differential equations can be recast in a compact form into the following general nonlinear Schr\"odinger equation: 
\begin{eqnarray}
i\hbar\partial_t\psi(q,t)&=&-\sum_{j=1}^N\frac{\hbar^2}{2m_j}\partial_{q_j}^2\psi(q,t)+V(q)\psi(q,t)\nonumber\\
&+&\mathcal{N}_f\big(|\psi(q)|^2\big)\psi(q,t), 
\label{nonlinear Schroedinger equation general}
\end{eqnarray} 
that is, Eqs. (\ref{continuity equation}) and (\ref{modified HJM equation}) are respectively the imaginary and the real parts of Eq. (\ref{nonlinear Schroedinger equation general}). Furthermore, in the limit of vanishing $\mathcal{N}_f$, we regain the standard linear Schr\"odinger equation
\begin{eqnarray}
i\hbar\partial_t\psi(q,t)&=&-\sum_{j=1}^N\frac{\hbar^2}{2m_j}\partial_{q_j}^2\psi(q,t)+V(q)\psi(q,t). 
\label{Schroedinger equation}
\end{eqnarray}
$\mathcal{N}_f$ defined in Eq. (\ref{correction to HJM and S equations}) thus determines the form and strength of the nonlinearity in the Schr\"odinger equation of Eq. (\ref{nonlinear Schroedinger equation general}). Finally, when the estimation error $\epsilon_{p_f}(q;\xi)$ is much smaller than the estimator $\partial_qS$, or the global fluctuation $\xi$ is ignorable, the third and fifth terms in Eq. (\ref{modified HJM equation}) (i.e., the $\hbar-$dependent terms) are ignorable, so that it reduces smoothly to the classical Hamilton-Jacobi equation: $\partial_tS+\sum_{j=1}^N\frac{(\partial_{q_j}S)^2}{2m_j}+V=0$. 

One can see that the above general scheme of estimation of momentum given positions under epistemic restriction provides a flexible framework to  construct a broad class of nonlinear variants of Schr\"odinger equation with a transparent operational meaning. As a concrete example, first, consider an estimation scheme so that $f$ that appears in the estimation error of Eq. (\ref{generalized estimation error}) has the following form:
\begin{eqnarray}
f_j\big(\rho_{\tilde{p}}(q)\big)=\Lambda_j\rho_{\tilde{p}}(q)^{\alpha}, 
\label{correction of estimation error: quadratic nonlinearity}
\end{eqnarray}
$j=1,\dots,N$, where $\Lambda_j$ is a real parameter with the dimension ${\rm [length]}^{-1}$, and $\alpha$ is a non-vanishing real number. In the limit $\Lambda_j\rightarrow 0$, we have $f_j\rightarrow 0$, $j=1,\dots,N$, so that the estimation error of Eq. (\ref{generalized estimation error}) reduces back to the specific form assumed in Sec. \ref{An epistemic reconstruction of quantum mechanics: estimation under epistemic restriction} given by Eq. (\ref{estimation error}). Inserting Eq. (\ref{correction of estimation error: quadratic nonlinearity}) into Eq. (\ref{average energy correction}), one has 
\begin{eqnarray}
 D_f\big[\rho_{\tilde{p}}\big]=\sum_{j=1}^N\int{\rm d}q\Big(\frac{\hbar^2\Lambda_j}{4m_j}\rho_{\tilde{p}}^{\alpha}\partial_{q_j}\rho_{\tilde{p}}+\frac{\hbar^2\Lambda_j^2}{8m_j}\rho_{\tilde{p}}^{2\alpha+1}\Big).
\label{average energy correction for quadratic nonlinearity}
\end{eqnarray}
From Eq. (\ref{correction to HJM and S equations}), we therefore obtain 
\begin{eqnarray}
\mathcal{N}_f\big(\rho_{\tilde{p}}(q)\big)=\Omega\rho_{\tilde{p}}(q)^{2\alpha}=\Omega|\psi(q)|^{4\alpha}, 
\label{polynomial nonlinearity}
\end{eqnarray}
where $\Omega=\sum_{j=1}^N\frac{\hbar^2\Lambda_j^2}{8m_j}(2\alpha+1)$, and we have used Eq. (\ref{Born's quadratic law}) in the last equality. Inserting into Eq. (\ref{nonlinear Schroedinger equation general}) we finally obtain the following polynomial nonlinear Schr\"odinger equation: 
\begin{eqnarray}
i\hbar\partial_t\psi(q,t)&=&-\sum_{j=1}^N\frac{\hbar^2}{2m_j}\partial_q^2\psi(q,t)+V(q)\psi(q,t)\nonumber\\
&+&\Omega|\psi(q,t)|^{4\alpha}\psi(q,t). 
\label{quadratic nonlinear Schroedinger equation}
\end{eqnarray} 
which reduces to the well-known quadratic nonlinear Schr\"odinger equation for $\alpha=1/2$. 

As another example, and for later comparison, consider an estimation scheme so that $f$ in Eq. (\ref{generalized estimation error}) has the following form:
\begin{eqnarray}
f_j\big(\rho_{\tilde{p}}(q),\partial_q\rho_{\tilde{p}}(q)\big)=\Lambda_j\Big(\frac{\partial_{q_j}\rho_{\tilde{p}}(q)}{\rho_{\tilde{p}}(q)}\Big)^{\beta},
\label{correction of estimation error: satisfying estimation independence}
\end{eqnarray}
$j=1,\dots,N$, where $\beta >1$, and $\Lambda_j$ is a real parameter with the dimension of [length]$^{\beta-1}$. We have thus assumed a higher order error term postulated in Eq. (\ref{estimation error}). Inserting Eq. (\ref{correction of estimation error: satisfying estimation independence}) into Eq. (\ref{average energy correction}), we obtain 
\begin{eqnarray}
&&D_f[\rho_{\tilde{p}}]\nonumber\\
&=&\sum_{j=1}^N\int{\rm d}q\Big(\frac{\hbar^2\Lambda_j}{4m_j}\partial_{q_j}\rho_{\tilde{p}}\Big(\frac{\partial_{q_j}\rho_{\tilde{p}}}{\rho_{\tilde{p}}}\Big)^{\beta}+\frac{\hbar^2\Lambda_j^2}{8m_j}\rho_{\tilde{p}}\Big(\frac{\partial_{q_j}\rho_{\tilde{p}}}{\rho_{\tilde{p}}}\Big)^{2\beta}\Big). 
\end{eqnarray}
Finally, using Eq. (\ref{correction to HJM and S equations}), the nonlinearity $\mathcal{N}_f$ in the Schr\"odinger equation of Eq. (\ref{nonlinear Schroedinger equation general}) can be computed to get 
\begin{eqnarray}
&&\mathcal{N}_f(|\psi|^2)\nonumber\\
&=&\sum_{j=1}^N\frac{\hbar^2\Lambda_j}{4m_j}\Big[-\beta\Big(\frac{\partial_{q_j}|\psi|^2}{|\psi|^2}\Big)^{\beta+1}-(\beta+1)\partial_{q_j}\Big(\frac{\partial_{q_j}|\psi|^2}{|\psi|^2}\Big)^{\beta}\Big]\nonumber\\
&+&\sum_{j=1}^N\frac{\hbar^2\Lambda_j^2}{8m_j}\Big[-(2\beta-1)\Big(\frac{\partial_{q_j}|\psi|^2}{|\psi|^2}\Big)^{2\beta}-{2\beta}\partial_{q_j}\Big(\frac{\partial_{q_j}|\psi|^2}{|\psi|^2}\Big)^{2\beta-1}\Big], 
\label{nonlinear Schroedinger equation satisfying EI}
\end{eqnarray}
where we have used Eq. (\ref{Born's quadratic law})

Let us give a few remarks concerning the derivation of the class of nonlinear variants of Schr\"odinger equation of Eq. (\ref{nonlinear Schroedinger equation general}). First, we note importantly that defining the wave function as in Eq. (\ref{wave function}) amounts to the assumption that the Born's quadratic law of Eq. (\ref{Born's quadratic law}) is kept valid. This is unlike the generalizations of quantum mechanics suggested in Refs. \cite{Valentini nonequilibrium,Aaronson nonlinearity-nonunitary - fast computation}, wherein the Born's quadratic law is somehow violated. 

Notice that the epistemic decomposition of the momentum field of Eq. (\ref{generalized fundamental epistemic decomposition}) is invariant under the addition of a global phase to the wave function, i.e., $\psi\mapsto e^{i\alpha}\psi$, where $\alpha$ is an arbitrary real number. But, unlike the specific estimation scheme of Sec. \ref{An epistemic reconstruction of quantum mechanics: estimation under epistemic restriction} with the epistemic decomposition of momentum field given by Eq. (\ref{fundamental epistemic decomposition}), that in Eq. (\ref{generalized fundamental epistemic decomposition}) is in general no longer invariant under the more general transformation of wave function: $\psi\mapsto Z\psi$, where $Z$ is an arbitrary complex number. We note however that while the epistemic decomposition of momentum field of Eq. (\ref{generalized fundamental epistemic decomposition}) with the specific $f$ given by Eq. (\ref{correction of estimation error: quadratic nonlinearity}) is not invariant under the transformation $\psi\mapsto Z\psi$, that with the specific $f$ given by Eq. (\ref{correction of estimation error: satisfying estimation independence}) is. As will be argued in Sec. \ref{discussion}, the two different $f$s in Eqs. (\ref{correction of estimation error: quadratic nonlinearity}) and (\ref{correction of estimation error: satisfying estimation independence}), leading to two different variants of nonlinearity in the Schr\"odinger equations respectively given by Eqs. (\ref{polynomial nonlinearity}) and (\ref{nonlinear Schroedinger equation satisfying EI}), are also fundamentally distinguished with respect to certain inferential-causality principle. One can also see that the form of the nonlinearity $\mathcal{N}_f$ determined in Eq. (\ref{correction to HJM and S equations}) does not depend on $S(q)$ which is due to the assumption that $f$ in Eq. (\ref{generalized estimation error}) does not depend on $S(q)$ either. Of course, it can be mathematically extended to depend also on $S(q)$. However, in this case, both the estimator of Eq. (\ref{weakly unbiased best estimator}) and the estimation error depend on $S(q)$, so that they are no longer independent of each other which is undesirable from the information theoretical point of view. 

Note further that, using the definition of wave function in Eq. (\ref{wave function}), the average energy given in Eq. (\ref{average energy}) can be written in terms of wave function as 
\begin{eqnarray}
\braket{H}_{\{S,\rho_{\tilde{p}}\}}=\braket{\psi|\hat{H}|\psi}+D_f[|\psi|^2], 
\label{general average energy in terms of wave function}
\end{eqnarray}
where $\hat{H}=\sum_{j=1}^N\hat{p}_j^2/2m_j+V(\hat{q})$ is the usual Hermitian quantum Hamiltonian. Hence, $D_f[|\psi|^2]$ defined in Eq. (\ref{average energy correction}) provides the correction to the quantum average energy $\braket{\psi|\hat{H}|\psi}$. The functional form of this correction term depends on the estimation error of Eq. (\ref{generalized estimation error}) via $f(\rho_{\tilde{p}},\partial_q\rho_{\tilde{p}})=f(|\psi|^2,\partial_q|\psi|^2)$, and is responsible for the appearance of the nonlinearity $\mathcal{N}_f$ in the Schr\"odinger equation of Eq. (\ref{nonlinear Schroedinger equation general}) via Eq. (\ref{correction to HJM and S equations}). The correction of average energy $D_f[|\psi|^2]$, and thus the nonlinearity $\mathcal{N}_f(|\psi|^2)$, vanishes for all $\psi$ (i.e., for all estimation schemes characterized by $(S,\rho_{\tilde{p}})$), iff $f=0$, so that the generalized estimation error of Eq. (\ref{generalized estimation error}) reduces back to the specific estimation error of Eq. (\ref{estimation error}) leading to the standard quantum mechanics. Moreover, unlike the quantum average energy $\braket{\psi|\hat{H}|\psi}$, the correction term $D_f[|\psi|^2]$ is in general not bi-linear in $\psi$. 

Next, as in the case of linear Schr\"odinger equation, the nonlinear Schr\"odinger equation of Eq. (\ref{nonlinear Schroedinger equation general}) conserves the average energy and probability current. In fact, as demonstrated above, we have upgraded the conservation of average energy and conservation of trajectories (which implies the conservation of probability current) as the principles which single out the dynamical equation when the agent does not make any selection of trajectories, encompassing both the linear and nonlinear variants of the Schr\"odinger equation \cite{Agung epistemic interpretation}  (see also Appendix \ref{A concrete illustration of the epistemic reconstruction of quantum mechanics within the operational scheme of estimation under epistemic restriction}). We emphasize that the above two constraints, i.e., conservation of trajectories and average energy, are not agent-independent objective physical constraint like the principle of least action. Rather, they are subjective epistemic constraints conditional on the agent's action that she does not make a selection of trajectories manifested in the setting of the experiment. Hence, the nonlinear Schr\"odinger equation of Eq. (\ref{nonlinear Schroedinger equation general}) should be seen as a Bayesian-inferential rule to update the agent's estimation about her system when she does not make any selection of trajectories. 

If the agent instead makes a selection trajectories, she must no longer impose conservation of average energy and trajectories, so that the Bayesian updating of her estimation no longer follows the nonlinear Schr\"odinger equation of Eq. (\ref{nonlinear Schroedinger equation general}). Such a selection of trajectories is necessary when the agent makes a measurement \cite{Agung epistemic interpretation}  (see also Appendix \ref{A concrete illustration of the epistemic reconstruction of quantum mechanics within the operational scheme of estimation under epistemic restriction}). From this observation, the nonlinearity in the Schr\"odinger equation of Eq. (\ref{nonlinear Schroedinger equation general}) therefore clearly, by construction, has nothing to do with the problem of Schr\"odinger's cat, unlike those nonlinearities discussed in Refs. \cite{Pearle nonlinearity,Gisin nonlinearity-stochasticity,GRW theory,Diosi gravity induced collapse,Bassi collapse model review} which were ad-hoc-ly introduced to circumvent this central aspect of quantum measurement problem. We note additionally that within our estimation scheme, since we assume that the system has a definite configuration all the time as in classical mechanics, by construction, there is no problem of Schr\"odinger's cat.   

Finally, we mention that some authors have proposed several different frameworks to introduce nonlinearities in the Schr\"odinger equation \cite{Bialynicki-Birula nonlinearity,Weinberg nonlinearity,Mielnik generalized quantum mechanics,Haag-Bannier nonlinear SE,Kibble nonlinear SE}, with the main goal to give a guide for stringent precision test of quantum mechanics. In particular, Weinberg  offered an elegant general `Hamiltonian framework' to nonlinearly generalize quantum mechanics \cite{Weinberg nonlinearity}. This is done by assuming that, unlike in standard quantum mechanics, the average energy, or, the `Hamiltonian functional', is in general non-bilinear in the wave functions as in our epistemic model. Moreover, the time evolution, i.e., the nonlinear Schr\"odinger equation is obtained by imposing the action principle. We emphasize that in Weinberg's approach, the nonlinearities are devised as possible mathematical innovations relative to the original linear theory, rather than motivated by deep conceptual reflection. While mathematically elegant and beautiful, the physical origin and operational meaning of the nonlinearities are not entirely clear so that the physical correspondence with the original linear theory is not conceptually transparent. Moreover, it suffers the same interpretational problem as that of the linear theory. 

By contrast, here we work within a general epistemic framework based on an operationally transparent scheme of estimation of momentum given the positions under epistemic restriction. Most importantly, unlike those in Refs. \cite{Bialynicki-Birula nonlinearity,Weinberg nonlinearity,Mielnik generalized quantum mechanics,Haag-Bannier nonlinear SE,Kibble nonlinear SE}, within the general scheme of estimation, by construction, the nonlinearity in the Schr\"odinger equations has a conceptually transparent operational meaning in terms of generalization of estimation errors. This transparent interpretation of the nonlinearities in the Schr\"odinger equation in terms of generalization of estimation errors, as will be discussed in Sec. \ref{discussion}, allows us to impose a physically transparent and reasonable inferential-causality principle which rules out a broad class of nonlinear generalizations of the Schr\"odinger equation. Another important conceptual advantage of our general epistemic framework based on the scheme of estimation of momentum given positions is that, as will be shown in Sec. \ref{A class of modified uncertainty relations}, we can directly derive the associated modifications of the Heisenberg uncertainty principle, and study its transparent relation with the resulting nonlinearities in the Schr\"odinger equation.

The above observation also suggests an interesting point that the principle of conservation of average energy and trajectories employed in the present manuscript to derive the (linear and nonlinear) Schr\"odinger equation are deeply connected with the action principle used in Weinberg's approach. A derivation of the (linear and nonlinear) Schr\"odinger equation using action principle, following that in Ref. \cite{Hall quantum-classical hybrid model} which is closely related to our derivation, is given in the Appendix \ref{Derivation of the nonlinear Schroedinger equation via action principle}. Note importantly however that unlike the least action principle which is objective independent of the agent's action, the principle of conservation of average energy and trajectories employed in the present work are epistemic or informational, conditional on the agent's action. Within our model, as discussed above, the Schr\"odinger equation arises only when the agent does not make measurement (i.e., she does not make a selection of trajectories associated with the measurement outcomes) so that the conservation of average energy and trajectories apply. By contrast, from the principle of least action, it seems to be unclear why (at least the linear) Schr\"odinger equation only applies when the agent does not make a measurement. Hence, while the two approaches lead to the same equation, the meaning of the resulting Schr\"odinger equation are different. Moreover, the principle of conservation of average energy and trajectories are natural, transparent and intuitive, whereas the principle of least action is somehow ad-hoc. 

\subsection{A class of generalized Heisenberg uncertainty principle \label{A class of modified uncertainty relations}}

In this section we derive a broad class of generalized Heisenberg uncertainty principle from the general estimation error of Eq. (\ref{generalized estimation error}). For notational simplicity, we consider a system with one spatial degree of freedom. Note before proceeding that to derive the uncertainty relations rigorously within the epistemic framework based on the generalized scheme of estimation, we need to develop a detailed mechanism of measurement. We shall however not pursue this problem, and instead assume that any reliable measurement mechanism within the generalized scheme of estimation must satisfy a reasonable informational requirement to be mentioned below. 

Consider first the estimation scheme discussed in Sec. \ref{An epistemic reconstruction of quantum mechanics: estimation under epistemic restriction}, namely when $f$ in Eq. (\ref{generalized estimation error}) is vanishing so that the estimation error takes the specific form given by Eq. (\ref{estimation error}). This specific estimation scheme, as elaborated in Refs. \cite{Agung epistemic interpretation,Agung-Daniel model}, reproduces the prediction of standard quantum mechanics. In this case, it was shown in Ref. \cite{Agung-Daniel model} that, in general, the ensemble average of a physical quantity $O(p,q)$ up to second order in $p$, is equal to the average of the outcomes of the quantum measurement of a Hermitian quantum observable $\hat{O}$ associated with $O$, i.e., 
\begin{eqnarray}
\braket{O}_{\{S,\rho_{\tilde{p}}\}}&=&\int{\rm d}q{\rm d}p{\rm d}\xi O(p,q){\rm P}_{\{S,\rho_{\tilde{p}}\}}(p,q|\xi)\chi(\xi)\nonumber\\
&=&\braket{\psi|\hat{O}|\psi}=\sum_jo_j{\rm P}(o_j|\psi),
\label{measurement outcomes as consistent estimator}
\end{eqnarray} 
where  we have used Eqs. (\ref{Planck constant}) and (\ref{wave function}). Here ${\rm P}_{\{S,\rho_{\tilde{p}}\}}(p,q|\xi)=\sum_{j=1}^N\delta\big(p_j-\tilde{p}_j(q;\xi)\big)\rho_{\tilde{p}}(q)$ with $\tilde{p}(q;\xi)$ defined in Eq. (\ref{fundamental epistemic decomposition}), $o_j$, $j=1,2,\dots$ is the eigenvalue of $\hat{O}$, and ${\rm P}(o_j|\psi)$ is the probability to obtain outcome $o_j$. This equality suggests that, while each single measurement outcome given by one of the eigenvalues of $\hat{O}$ does not in general reveal the objective value of $O$ prior to measurement, each single measurement outcome can be seen as an unbiased estimate of the average value of $O$, so that the average of the measurement outcomes is equal to the average of $O$ as expressed in Eq. (\ref{measurement outcomes as consistent estimator}). In particular, as a corollary of Eq. (\ref{measurement outcomes as consistent estimator}), we have $\sigma_{\hat{p}}^2\doteq\braket{\psi|(\hat{p}-\braket{\psi|\hat{p}|\psi})^2|\psi}=\braket{(p-\braket{p}_{\{S,\rho_{\tilde{p}}\}})^2}_{\{S,\rho_{\tilde{p}}\}}=\sigma_p^2$, and similarly $\sigma_{\hat{q}}^2\doteq\braket{\psi|(\hat{q}-\braket{\psi|\hat{q}|\psi})^2|\psi}=\braket{(q-\braket{q}_{\{S,\rho_{\tilde{p}}\}})^2}_{\{S,\rho_{\tilde{p}}\}}\doteq\sigma_q^2$. Namely, the variance of the outcomes of momentum (position) measurement, i.e., $\sigma_{\hat{p}}^2 (\sigma_{\hat{q}}^2)$, is equal to the variance of the  momentum (position) of the statistical model, $\sigma_p^2(\sigma_q^2)$. 

We assume below that the above conclusion drawn from the case when $f=0$ | namely that the statistical mean of measurement outcomes of physical quantities (up to second order in momentum) reproduces the statistical mean of the corresponding (classical) physical quantity of the underlying statistical model | can be carried over to the case when $f\neq 0$. A similar assumption is also postulated by Weinberg in his general Hamiltonian framework for introducing nonlinearity in the Schr\"odinger equation \cite{Weinberg nonlinearity}. Hence, we require that, within the epistemic framework based on the generalized estimation scheme with the estimation error given by Eq. (\ref{generalized estimation error}), even when $f\neq 0$, any reliable measurement scheme must be such that the variance of the outcome of the momentum measurement is equal to the variance of the momentum $p$ of the underlying statistical model, denoted by $\sigma_{p_f}^2$ (with a subscript $f$). Similarly, the variance of the outcome of the position measurement is equal to the variance of $q$ of the statistical model $\sigma_{q_f}^2$. To study the uncertainty relation between the statistics of the outcomes of measurement of momentum and position in this generalized estimation scheme, it is thus sufficient to develop the uncertainty relation between $\sigma_{p_f}^2$ and $\sigma_{q_f}^2$. 

First, from Eq. (\ref{generalized estimation error}), we can compute the MS error for the estimation of momentum field to obtain, noting Eq. (\ref{Planck constant}),
\begin{eqnarray}
\mathcal{E}_{p_f}^2=\int{\rm d}q(\epsilon_{p_f}(q;\xi))^2\chi(\xi)\rho_{\tilde{p}}(q)=\frac{\hbar^2}{4}J_{q_f}+C_f.
\label{information trade-off modified estimation scheme}
\end{eqnarray}
Here, $J_{q_f}\doteq\int{\rm d}q\big(\frac{\partial_q\rho_{\tilde{p}}(q)}{\rho_{\tilde{p}}(q)}\big)^2\rho_{\tilde{p}}(q)$ is the Fisher information about the mean position contained in $\rho_{\tilde{p}}(q)$, and $C_f$ is a functional of $\rho_{\tilde{p}}(q)$ defined as 
\begin{eqnarray}
C_f[\rho_{\tilde{p}}]&\doteq&\frac{\hbar^2}{4}\int{\rm d}q\Big(2\frac{\partial_q\rho_{\tilde{p}}}{\rho_{\tilde{p}}}f+f^2\Big)\rho_{\tilde{p}}(q)\nonumber\\
&=&2m D_f\big[\rho_{\tilde{p}}\big], 
\label{the strength of correction of UR}
\end{eqnarray}
where we have used Eq. (\ref{average energy correction}) in the second equality which is valid for the specific case of particles in a scalar potential. On the other hand, in the estimation of mean position $q_o\doteq\int{\rm d}q q\rho_{\tilde{p}}(q)$ with the unbiased estimator $q$, the associated MS error must satisfy the Cram\'er-Rao inequality \cite{Papoulis and Pillai book on probability and statistics}:
\begin{eqnarray}
\mathcal{E}_{q_f}^2=\int{\rm d}q(q-q_o)^2\rho_{\tilde{p}}(q)\ge\frac{1}{J_{q_f}}.
\label{Cramer-Rao inequality for estimation of mean position}
\end{eqnarray}
Combining Eq. (\ref{information trade-off modified estimation scheme}) with Eq. (\ref{Cramer-Rao inequality for estimation of mean position}), we thus obtain the following uncertainty relation between the MS errors of the simultaneous estimation of momentum field and mean position:
\begin{eqnarray}
\mathcal{E}_{p_f}^2\mathcal{E}_{q_f}^2\ge\frac{\hbar^2}{4}+\frac{C_f}{J_{q_f}}. 
\label{MS errors trade-off modified estimation scheme}
\end{eqnarray}

On the other hand, from Eq. (\ref{generalized fundamental epistemic decomposition}), the variance of the momentum can be computed to obtain 
\begin{eqnarray}
\sigma_{p_f}^2=\Delta_{p_f}^2+\mathcal{E}_{p_f}^2, 
\label{decomposition of variance of momentum as dispersion and accuracy of estimation: modified UR}
\end{eqnarray}
where we have used Eq. (\ref{Planck constant}), $\mathcal{E}_{p_f}^2$ is given in (\ref{information trade-off modified estimation scheme}), and $\Delta_{p_f}^2\doteq\int{\rm d}q\big(\partial_qS(q)-\int{\rm d}q'\partial_{q'}S(q')\rho_{\tilde{p}}(q')\big)^2\rho_{\tilde{p}}(q)$ is the variance of the estimator $\partial_qS(q)$. Hence, the variance of the momentum can be decomposed into the accuracy of the estimation of momentum $\mathcal{E}_{p_f}^2$ of Eq. (\ref{information trade-off modified estimation scheme}), and the precision of the estimation $\Delta_{p_f}^2$. Moreover, one also straightforwardly has $\sigma_{q_f}^2=\mathcal{E}_{q_f}^2$. Multiplying this with Eq. (\ref{decomposition of variance of momentum as dispersion and accuracy of estimation: modified UR}), and using Eq. (\ref{MS errors trade-off modified estimation scheme}), one finally obtains the following uncertainty relation between the variances of momentum and position:
\begin{eqnarray}
\sigma_{p_f}^2\sigma_{q_f}^2&=&\Delta_{p_f}^2\mathcal{E}_{q_f}^2+\mathcal{E}_{p_f}^2\mathcal{E}_{q_f}^2\nonumber\\
&\ge&\Delta_{p_f}^2\mathcal{E}_{q_f}^2+\frac{\hbar^2}{4}+\frac{C_f}{J_{q_f}}.
\label{modified Heisenberg-Kennard UR}
\end{eqnarray}
Furthermore, when $C_f=0$ we regain the Heisenberg-Kennard uncertainty relation \cite{Heisenberg UR,Kennard UR}
\begin{eqnarray}
\sigma_p^2\sigma_q^2\ge\Delta_p^2\mathcal{E}_q^2+\frac{\hbar^2}{4}\ge\frac{\hbar^2}{4},
\label{Heisenberg-Kennard UR}
\end{eqnarray}
where, e.g., $\sigma_p^2\doteq\sigma_{p_f}^2|_{C_f=0}$, et cetera. This is the case for all estimation schemes characterized by the pairs of $(S,\rho_{\tilde{p}})$, iff $f=0$ so that the estimation error of Eq. (\ref{generalized estimation error}) reduces back to the specific form given by Eq. (\ref{estimation error}). The last term on the right hand side of Eq. (\ref{modified Heisenberg-Kennard UR}) thus provides a nontrivial correction to the Heisenberg-Kennard uncertainty relation of Eq. (\ref{Heisenberg-Kennard UR}). In Ref. \cite{Agung estimation independence} we have derived Eq. (\ref{modified Heisenberg-Kennard UR}) but for a specific case of Eq. (\ref{generalized estimation error}) with $f=\Lambda\partial_q\rho_{\tilde{p}}$. 

Now, let us consider a specific preparation characterized by a Gaussian wave function, $\psi(q)=(2\pi\sigma_{q_f}^2)^{-1/4}e^{-(q-q_o)^2/4\sigma_{q_f}^2+ip_oq/\hbar}$. In this case, we have $\rho_{\tilde{p}}(q)=(2\pi\sigma_{q_f}^2)^{-1/2}e^{-(q-q_o)^2/2\sigma_{q_f}^2}$ so that $J_{q_f}=1/\sigma_{q_f}^2=1/\mathcal{E}_{q_f}^2$, and therefore Eq. (\ref{Cramer-Rao inequality for estimation of mean position}) is saturated. Noting Eq. (\ref{information trade-off modified estimation scheme}), it follows then that Eq. (\ref{MS errors trade-off modified estimation scheme}) is also saturated. Moreover, since for Gaussian wave function $S(q)=p_oq$, we have $\Delta_{p_f}^2=0$, Eq. (\ref{decomposition of variance of momentum as dispersion and accuracy of estimation: modified UR}) becomes $\sigma_{p_f}^2=\mathcal{E}_{p_f}^2$. Combining all these facts, we thus finally obtain, for Gaussian wave functions, 
\begin{eqnarray}
\sigma_{p_f}^2\sigma_{q_f}^2=\frac{\hbar^2}{4}+\sigma_{q_f}^2C_f, 
\label{generalized HK UR for Gaussian}
\end{eqnarray}
which reduces to the usual relation for Gaussian wave function in standard quantum mechanics when $C_f=0$, i.e., when $f=0$. Hence, for nonvanishing $C_f$, unlike in standard quantum mechanics, the product $\sigma_{p_f}^2\sigma_{q_f}^2$ of the variances of momentum and position depends on the profile of the Gaussian wave function, i.e., it is no longer invariant for all Gaussians. As a concrete example, consider the case when  $f$ is given by Eq. (\ref{correction of estimation error: quadratic nonlinearity}) with $\alpha=1/2$, so that $C_f$ in Eq. (\ref{generalized HK UR for Gaussian}) has the form $C_f=\frac{\hbar^2\Lambda^2}{4}\int{\rm d}q\rho_{\tilde{p}}^2=\frac{\hbar^2\Lambda^2}{8\pi^{1/2}\sigma_{q_f}}\ge 0$, where $\Omega=\frac{\hbar^2\Lambda^2}{4m}$ characterizes the strength of the nonlinearity in the quadratic nonlinear Schr\"odinger equation of Eq. (\ref{quadratic nonlinear Schroedinger equation}). Hence, in this case, we have $\sigma_{p_f}^2\sigma_{q_f}^2\ge\sigma_p^2\sigma_q^2=\frac{\hbar^2}{4}$, i.e., the model has a `stronger than quantum uncertainty'. Moreover, increasing the strength of the estimation error $\Lambda$, increases both the uncertainty and nonlinearity. Next let us consider the case when  $f$ is given by Eq. (\ref{correction of estimation error: satisfying estimation independence}) with $\beta=3$. In this case, we have $C_f=\frac{\hbar^2}{4\sigma_{q_f}^6}(6\Lambda\sigma_{q_f}^2+15\Lambda^2)$, so that $C_f< 0$ for $ -\frac{6}{15}\sigma_{q_f}^2<\Lambda< 0$, and $C_f\ge0$ otherwise. When $C_f <0$, we thus have $\sigma_{p_f}^2\sigma_{q_f}^2\le\sigma_p^2\sigma_q^2=\frac{\hbar^2}{4}$. Note that such a `weaker than quantum uncertainty' does not necessarily mean that the statistical model is more classical than quantum mechanics. This can be seen from the fact that even in this case $f$ in Eq. (\ref{correction of estimation error: satisfying estimation independence}) could be very large. 

\section{Discussion: nonlinearity, deviation from Heisenberg uncertainty, and estimation independence\label{discussion}}

We have shown that within the scheme of estimation of momentum given positions, with the estimator of Eq. (\ref{weakly unbiased best estimator}) and the generalized estimation errors of Eq. (\ref{generalized estimation error}) encapsulated (up to $\xi$) by the wave function defined in Eq. (\ref{wave function}), the agent's estimation when she does not make measurement, must be updated in time according the nonlinear Schr\"odinger equation of Eq. (\ref{nonlinear Schroedinger equation general}). Moreover, the variances of the outcomes of momentum and position measurements must satisfy the generalized Heisenberg-Kennard uncertainty relation of Eq. (\ref{modified Heisenberg-Kennard UR}). We emphasize that, by construction, both the nonlinearity in the Schr\"odinger equation and the deviation from the Heisenberg uncertainty principle, respectively characterized by  $\mathcal{N}_f$ and $C_f$ defined in Eqs. (\ref{correction to HJM and S equations}) and (\ref{the strength of correction of UR}), arise from the same estimation error of Eq. (\ref{generalized estimation error}) via $f$. They therefore should be closely related to each other. 

Indeed, in general, by construction, it is easy to see that no nonlinearity in the Schr\"odinger equation arises without a deviation from the Heisenberg uncertainty principle. In particular, noting Eqs. (\ref{correction to HJM and S equations}) and (\ref{the strength of correction of UR}), for a single one-dimensional particle of mass $m$, they are directly related as  
\begin{eqnarray}
\mathcal{N}_f\big(|\psi(q)|^2\big)=\frac{1}{2m}\frac{\delta C_f}{\delta\rho_{\tilde{p}}(q)}\Big|_{\rho_{\tilde{p}}(q)=|\psi(q)|^2}. 
\label{strength nonlinearity vs strength of violation of UR}
\end{eqnarray}
The above relation shows that to have nonlinearity in the Schr\"odinger equation for a single particle, the deviation from the Heisenberg uncertainty principle, i.e., $C_f$ defined in Eq. (\ref{the strength of correction of UR}), cannot be a functional linear in $\rho_{\tilde{p}}(q)$ or/and in its spatial derivatives, $\partial_q\rho_{\tilde{p}}(q)$. One may thus conclude that, within the estimation scheme, the nonlinearity is generated by the deviation from the Heisenberg uncertainty principle. Furthermore, from Eq. (\ref{strength nonlinearity vs strength of violation of UR}), since it is possible to have $C_f\neq 0$ with $\mathcal{N}_f=0$, one can still have a nontrivial deviation from the Heisenberg uncertainty relation without inducing nonlinearity in the Schr\"odinger equation. This is the case, for example, when $f=F(q)$, where $F$ is independent of $\rho_{\tilde{p}}(q)$ and $\partial_q\rho_{\tilde{p}}(q)$, so that from Eq. (\ref{the strength of correction of UR}), $\mathcal{C}_f$ is linear in $\rho_{\tilde{p}}(q)$ and $\partial_q\rho_{\tilde{p}}(q)$. It suggests that one can still have a superposition principle while the Heisenberg uncertainty relation is to some extent modified. 

Remarkably, within the epistemic reconstruction based on scheme of estimation of momentum given positions with the estimation error having the general form of Eq. (\ref{generalized estimation error}), noting Eqs.  (\ref{correction to HJM and S equations}) and (\ref{the strength of correction of UR}), and assuming that the definition of wave function is given by Eq. (\ref{wave function}), i.e., assuming that the Born's quadratic law of Eq. (\ref{Born's quadratic law}) stays solid, the linear Schr\"odinger equation of Eq. (\ref{Schroedinger equation}), and the exact form of Heisenberg-Kennard uncertainty relation of Eq. (\ref{Heisenberg-Kennard UR}), are regained iff $f=0$, so that the estimation error of Eq. (\ref{generalized estimation error}) reduces back to the specific form given by Eq. (\ref{estimation error}). Keeping this observation in mind, we may therefore conclude that not only standard quantum mechanics corresponds to a specific estimation scheme with the specific estimator and estimation error given respectively by Eqs. (\ref{weakly unbiased best estimator}) and (\ref{estimation error}), it is also difficult to nontrivially modify a part of quantum mechanics, e.g., the linearity of the Schr\"odinger equation, without changing the other fundamental parts of the theory, e.g., the exact form of the Heisenberg uncertainty principle. 

Finally, having obtained the various variants of Schr\"odinger equations given in Eq. (\ref{nonlinear Schroedinger equation general}) and uncertainty relations of Eq. (\ref{modified Heisenberg-Kennard UR}), how do we choose among them? To this end, remember first that the standard linear Schr\"odinger equation of Eq. (\ref{Schroedinger equation}) and the Heisenberg-Kennard uncertainty relation of Eq. (\ref{Heisenberg-Kennard UR}) have passed all stringent tests conceived to date. Moreover, there are striking theoretical results which suggest that nonlinearities in the Schr\"odinger equation and/or deviations from the exact Heisenberg uncertainty principle, may imply violations of some forms of causality, e.g., superluminal signalling \cite{Gisin nonlinearity - signaling,Polchinski nonlinearity - signaling,Czachor nonlinearity - signaling,Mielnik nonlinearity - signaling,Simon no-signaling imply linearity} and/or the second law of thermodynamics \cite{Peres nonlinearity violates 2nd law,Hanggi a deviation from UR violates 2nd}. It is therefore instructive to see, within the general epistemic framework based on the operational scheme of estimation of momentum given positions, if the specific estimation error given by Eq. (\ref{estimation error}), which together with the estimator of Eq. (\ref{weakly unbiased best estimator}) leads to the standard linear Schr\"odinger equation and the exact form of Heisenberg-Kennard uncertainty relation \cite{Agung epistemic interpretation,Agung-Daniel model}, might be justified based on some reasonable premises about causality. 

To investigate this last tantalizing question, let us discuss a physically transparent and plausible inferential-causality principle of estimation independence introduced in Ref. \cite{Agung estimation independence}. Consider two systems, referred to as system 1 and system 2, with a configuration $(q_1,q_2)$ and the corresponding conjugate momentum $(p_1,p_2)$, prepared independently of each other. First, recall that in classical mechanics, for such independent preparations of two systems, the total Lagrangian is decomposable, so that the associated Hamilton's principal function is also decomposable, i.e., $S_{\rm C}(q_1,q_2,t)=\int^{(q,t)}{\rm d}t'\big(L_1(q'_1,\dot{q}'_1)+L_2(q'_2,\dot{q}'_2)\big)=\int^{(q_1,t)}{\rm d}t'L_1(q'_1,\dot{q}'_1)+\int^{(q_2,t)}{\rm d}t'L_2(q'_2,\dot{q}'_2)=S_{{\rm C}_1}(q_1,t)+S_{{\rm C}_2}(q_2,t)$, where $L_j$ is the (classical) Lagrangian associated with system $j$, $j=1,2$. To have a smooth classical correspondence, it is therefore reasonable to assume that, within the generalized estimation scheme, $S(q)$ defined in Eq. (\ref{weakly unbiased best estimator}) for such pairs of independent preparations should also be decomposable: 
\begin{eqnarray}
S(q_1,q_2)=S_1(q_1)+S_2(q_2). 
\label{decomposable phase}
\end{eqnarray} 
Moreover, it is also natural to assume that in such pairs of independent preparations, the probability distribution of positions are factorizable, as in classical mechanics, i.e., 
\begin{eqnarray}
\rho_{\tilde{p}}(q_1,q_2)=\rho_{\tilde{p}_1}(q_1)\rho_{\tilde{p}_2}(q_2). 
\label{separable amplitude}
\end{eqnarray} 
Recalling the definition of wave function given in Eq. (\ref{wave function}), the above two assumptions amount to the postulate in standard quantum mechanics that the wave function associated with the independent preparations of the two systems is factorizable (unentangled), i.e., $\psi(q_1,q_2)=\sqrt{\rho_{\tilde{p}}}e^{iS/\hbar}=\sqrt{\rho_{\tilde{p}_1}\rho_{\tilde{p}_2}}e^{i(S_1+S_2)/\hbar}=\psi_1(q_1)\psi_2(q_2)$. The principle of estimation independence then requires that in such independent preparations, the estimation of momentum $\tilde{p}_j$ of system $j$, i.e., the associated estimator $\overline{p}_j$ and estimation error $\epsilon_{p_j}$, should be reasonably independent of the position $q_i$ of the system $i$, $i\neq j$, $i,j=1,2$ \cite{Agung estimation independence}. It thus captures an intuitive form of inferential-causality constraint. 

We shall impose the above plausible requirement to scrutinize the various estimation schemes discussed in the previous sections. Let us first consider the estimation scheme discussed in Sec. \ref{An epistemic reconstruction of quantum mechanics: estimation under epistemic restriction}, i.e., when the estimator and the estimation error take the specific forms respectively given by Eqs. (\ref{weakly unbiased best estimator}) and (\ref{estimation error}), leading to the standard quantum mechanics \cite{Agung epistemic interpretation,Agung-Daniel model}. Inserting Eq. (\ref{decomposable phase}) into Eq. (\ref{weakly unbiased best estimator}), one has 
\begin{eqnarray}
\overline{p}_j=\partial_{q_j}S(q_1,q_2)=\partial_{q_j}S_j(q_j),
\end{eqnarray}
$j=1,2$. Hence, the estimator $\overline{p}_j$ for estimating the momentum field $\tilde{p}_j$ of system $j$ is indeed independent of the position $q_i$ of system $i$, $i\neq j$, $i,j=1,2$, respecting the principle of estimation independence. Moreover, inserting Eq. (\ref{separable amplitude}) into Eq. (\ref{estimation error}), one obtains 
\begin{eqnarray}
\epsilon_{p_j}=\frac{\xi}{2}\partial_{q_j}\ln\rho_{\tilde{p}}(q_1,q_2)=\frac{\xi}{2}\partial_{q_j}\ln\rho_{\tilde{p}_j}(q_j),
\end{eqnarray}
$j=1,2$. Namely, the error $\epsilon_{p_j}$ of estimating $\tilde{p}_j$ of system $j$ is also independent of $q_i$ of system $i$, $i\neq j$, $i,j=1,2$, satisfying the requirement of estimation independence. In this sense, standard quantum mechanics with the linear Schr\"odinger equation and the exact form of the Heisenberg uncertainty principle reformulated within the operational scheme of estimation, thus elegantly respects the natural inferential-causality principle of estimation independence.  

Next, let us show that the above natural requirement of estimation independence is not fulfilled by a broad class of schemes of estimation of momentum given positions discussed in Sec. \ref{Nonlinear Schroedinger equation and modified Heisenberg uncertainty principle within an estimation scheme} with an estimator having the same form as that in Sec. \ref{An epistemic reconstruction of quantum mechanics: estimation under epistemic restriction} given by Eq. (\ref{weakly unbiased best estimator}), but with an estimation error of the form given by Eq. (\ref{generalized estimation error}) which generalizes Eq. (\ref{estimation error}) via a nonvanishing $f$. We only need to check whether the estimation error of Eq. (\ref{generalized estimation error}) passes the requirement of estimation independence. Since the first term on the right-hand side of Eq. (\ref{generalized estimation error}) is already shown above respecting the principle of estimation independence, we need only to examine the correction term $f$ under the estimation indepedence. 

Consider first the specific scheme of estimation of momentum given positions with the estimation error having the form of Eq. (\ref{generalized estimation error}) where $f$ is given by Eq. (\ref{correction of estimation error: quadratic nonlinearity}), leading to the polynomial nonlinear Schr\"odinger equation of Eq. (\ref{quadratic nonlinear Schroedinger equation}). Inserting Eq. (\ref{separable amplitude}) into Eq. (\ref{correction of estimation error: quadratic nonlinearity}), one has 
\begin{eqnarray}
&&f_j\big(\rho_{\tilde{p}}(q_1,q_2)\big)=f_j\big(\rho_{\tilde{p}_1}(q_1)\rho_{\tilde{p}_2}(q_2)\big)\nonumber\\
&=&\Lambda_j\big(\rho_{\tilde{p}_1}\rho_{\tilde{p}_2})^{\alpha}\neq\Lambda_j\rho_{\tilde{p}_j}^{\alpha}=f_j\big(\rho_{\tilde{p}_j}(q_j)\big), 
\label{violation of estimation independence for quadratic nonlinearity}
\end{eqnarray}
$j=1,2$. Hence, in this case, the error $\epsilon_{p_j}$ of estimating the momentum $\tilde{p}_j$ of system $j$ depends on the position $q_i$ of system $i$, $i\neq j$, $i,j=1,2$, even when the two systems are prepared independently of each other, violating the principle of estimation independence. In fact, one can check that any $f$ which is an analytical function only of $\rho_{\tilde{p}}$ (hence, independent of its spatial gradient) will not pass the reasonable requirement of estimation independence. This shows that, within the operational scheme of estimation of momentum given positions with the estimation error having the general form of Eq. (\ref{generalized estimation error}), the requirement of estimation independence rules out a broad class of forms of estimation errors, thus excludes a broad class of nonlinear generalizations of Schr\"odinger equation.   

By contrast, one can straightforwardly show that the estimation error of Eq. (\ref{generalized estimation error}) with $f$ given by Eq. (\ref{correction of estimation error: satisfying estimation independence}) satisfies the plausible requirement of estimation independence, i.e., inserting  Eq. (\ref{separable amplitude}) into Eq. (\ref{correction of estimation error: satisfying estimation independence}), we have
\begin{eqnarray}
f_j\big(\rho_{\tilde{p}_1}(q_1)\rho_{\tilde{p}_2}(q_2)\big)&=&\Lambda_j\Big(\frac{\partial_{q_j}\big(\rho_{\tilde{p}_1}(q_1)\rho_{\tilde{p}_2}(q_2)\big)}{\rho_{\tilde{p}_1}(q_1)\rho_{\tilde{p}_2}(q_2)}\Big)^{\beta}\nonumber\\
&=&\Lambda_j\Big(\frac{\partial_{q_j}\rho_{\tilde{p}_j}(q_j)}{\rho_{\tilde{p}_j}(q_j)}\Big)^{\beta}=f_j\big(\rho_{\tilde{p}_j}(q_j)\big), 
\end{eqnarray}
$j=1,2$. Indeed, all $f$ which has the form $f_j=G\big(\frac{\partial_{q_j}\rho_{\tilde{p}}}{\rho_{\tilde{p}}}\big)$ where $G$ is some scalar function of $\frac{\partial_{q_j}\rho_{\tilde{p}}}{\rho_{\tilde{p}}}$, satisfies the requirement of estimation independence. Note however that while this class of forms of $f$ does fulfil the requirement of estimation independence, it in general does not transform in the same way as the rest of terms in the epistemic decomposition of momentum field of Eq. (\ref{generalized fundamental epistemic decomposition}), so that the latter does not transform covariantly. 

A different kind of $f$ which satisfies the requirement of estimation independence takes the form $f_j=G\big(\frac{\partial_{q_j}\rho_{\tilde{p}}}{\rho_{\tilde{p}}},\partial_{q_j}S\big)$, $j=1,\dots,N$. This can be checked directly for two systems prepared independently of each other so that Eqs. (\ref{decomposable phase}) and (\ref{separable amplitude}) apply. Namely, we have: $f_j(\rho_{\tilde{p}_1}\rho_{\tilde{p}_2},S_1+S_2)=G\Big(\frac{\partial_{q_j}(\rho_{\tilde{p}_1}\rho_{\tilde{p}_2})}{\rho_{\tilde{p}_1}\rho_{\tilde{p}_2}},\partial_{q_j}(S_1+S_2)\Big)=G\Big(\frac{\partial_{q_j}\rho_{\tilde{p}_j}}{\rho_{\tilde{p}_j}},\partial_{q_j}S_j\Big)=f_j(\rho_{\tilde{p}_j},S_j)$, $j=1,2$. One can work out directly that such a choice of $f$ will lead to a different class of nonlinear variants of Schr\"odinger equation and generalized Heisenberg uncertainty principle. Note however that in this case, the estimation error becomes correlated with the estimator $\partial_qS$ which is unappealing from the information theoretical view point. 

Notice that when $f$ satisfies the requirement of estimation independence, e.g., that given by Eq. (\ref{correction of estimation error: satisfying estimation independence}), the associated correction term $D_f[\rho_{\tilde{p}}]$ to the quantum average energy defined in Eq. (\ref{average energy correction}) for two non-interacting systems is decomposable into that of each system. This can be seen directly by inserting Eq. (\ref{separable amplitude}) into Eq. (\ref{average energy correction}) for such $f$s. Accordingly, in this case, the nonlinearity $\mathcal{N}_f$ defined in Eq. (\ref{correction to HJM and S equations}) for two non-interacting systems is also decomposable, i.e., one has 
\begin{eqnarray}
\mathcal{N}_f(\rho_{\tilde{p}_1}(q_1)\rho_{\tilde{p}_2}(q_2))=\mathcal{N}_f(\rho_{\tilde{p}_1}(q_1))+\mathcal{N}_f(\rho_{\tilde{p}_2}(q_2)), 
\label{separability condition from estimation independence}
\end{eqnarray}
as is exemplified by the nonlinearity in Eq. (\ref{nonlinear Schroedinger equation satisfying EI}). This is not the case when $f$ does not respect the principle of estimation independence, as e.g., that given by Eq. (\ref{correction of estimation error: quadratic nonlinearity}) with the associated nondecomposable nonlinearity in Eq. (\ref{polynomial nonlinearity}). Within the estimation scheme, the principle of estimation independence thus implies that the product of two wave functions associated with two non-interacting systems, will evolve in time independently of each other, as intuitively expected. Such a natural separability condition for the dynamics of non-interacting systems is employed to single out the logarithmic nonlinear Schr\"odinger equation by Bialynicki-Birula and Mycielski \cite{Bialynicki-Birula nonlinearity}. Moreover, the separability for the dynamics of non-interacting systems are attained in Weinberg's Hamiltonian formalism by imposing the Homogeneity condition together with the additivity of the Hamiltonian functional \cite{Weinberg nonlinearity}. We emphasize that within our estimation scheme, unlike the latter two approaches, the separability condition for the dynamics of non-interacting systems has a transparent operational interpretation in terms of a natural inferential-causality principle of estimation independence. 

Next, it is interesting to note that, in the estimation scheme with the generalized estimation error of Eq. (\ref{generalized estimation error}), and $f$ is given by Eq. (\ref{correction of estimation error: quadratic nonlinearity}) which does not comply with the principle of estimation independence, the associated epistemic decomposition of the momentum fields of Eq. (\ref{generalized fundamental epistemic decomposition}) is not invariant under the transformation of wave function $\psi\rightarrow Z\psi$, where $Z$ is an arbitrary complex number. In contrast to this, for $f$ given by Eq. (\ref{correction of estimation error: satisfying estimation independence}) which complies with the estimation independence, the associated epistemic decomposition of the momentum fields of Eq. (\ref{generalized fundamental epistemic decomposition}) is invariant under the transformation $\psi\rightarrow Z\psi$. It is instructive to ask if this nice relation between the principle of estimation independence and the invariance of epistemic decomposition of the momentum fields of Eq. (\ref{generalized fundamental epistemic decomposition}) under $\psi\rightarrow Z\psi$ applies for all forms of $f$. Since the invariance of the nonlinear Schr\"odinger equation with respect to $\psi\rightarrow Z\psi$ is obtained in the Weinberg's Hamiltonian formalism by imposing the homogeneity condition to the Hamiltonian functional \cite{Weinberg nonlinearity}, this suggests a possible deep connection between the mathematical condition of homogeneity and the physically transparent inferential-causality principle of estimation independence, worth further study in the future. 

All the above observations show that the plausible inferential-causality principle of estimation independence puts a tight physical-informational constraint which rules out a significantly large class of mathematically possible modifications of standard quantum mechanics. Indeed, we have argued in Ref. \cite{Agung estimation independence} that, requiring the estimation error $\epsilon_p(q;\xi)$ for estimating the momentum given positions to satisfy the following conditions: (i) independent of the estimator $\overline{p}=\partial_qS$, (ii) transforms covariantly with the estimator, and (iii) respecting the principle of estimation independence, will single out the specific form of estimation error given by Eq. (\ref{estimation error}) up to the statistics of $\xi$, which has been argued in Refs. \cite{Agung epistemic interpretation,Agung-Daniel model} to imply the standard quantum mechanics. We note that Simon et al. in Ref. \cite{Simon no-signaling imply linearity} argued that the principle of no-signaling can be used to single out the linear quantum dynamics, by assuming, at the outset, the quantum kinematics and the quantum trace rule for computing the probability of measurement outcomes. See also Ref. \cite{Mielnik nonlinearity - signaling} for a similar argument. By contrast, within the above estimation scheme, the principle of estimation independence is used to reconstruct the underlying quantum kinematics by constraining the allowed forms of estimation error, without assuming any quantum structures. While we have assumed the Born's quadratic law of Eq. (\ref{Born's quadratic law}) via the definition of wave function in Eq. (\ref{wave function}), it is not the same as, and weaker than, assuming the quantum trace rule as in Ref. \cite{Simon no-signaling imply linearity}. Moreover, within the epistemic reconstruction framework based on the operational scheme of estimation, the linear Schr\"odinger equation follows from the kinematics via imposing the conservation laws, i.e., the conservation of trajectories and average energy, naturally embodying the assumption that the agent does not make any measurement via a selection of trajectories.  

\section{Conclusions and Remarks \label{conclusion}}

We have generalized the specific operational scheme of estimation of momentum given positions under epistemic restriction to reconstruct quantum mechanics proposed in Refs. \cite{Agung epistemic interpretation,Agung-Daniel model}, by considering a more general class of estimation errors. We showed that, provided Born's quadratic law is kept intact, it leads to a broad class of nonlinear variants of Schr\"odinger equation when the agent does not make measurement, and a class of generalized Heisenberg uncertainty principle. Within the operational scheme of estimation, both the nonlinearities in the Schr\"odinger equation and the deviation from the Heisenberg uncertainty principle have thus a transparent operational interpretation in terms generalization of the estimation errors. Hence, they are deeply related to each other; in particular, no nonlinearity in the Schr\"odinger equation without a deviation from the Heisenberg uncertainty principle. With this in mind, it is interesting to further study the connection between the deviation from the Heinseberg uncertainty principle which allows stronger than quantum correlation \cite{Ver Steeg relaxing uncertainty relation,Oppenheim-Wehner entropic UR and QS} and in turn may imply implausible computational power \cite{Popescu review,Dam informational approach Tsirelson bound,Brassard informational approach Tsirelson bound,Buhrman superstrong cryptography,Linden nonlocal computation,Brunner trivial communication,Pawlowski informational approach Tsirelson bound,Gross trivial dynamics with superstrong correlation},  and the nonlinearity in the  Schr\"odinger equation which may lead to a violation of no-signaling \cite{Gisin nonlinearity - signaling,Polchinski nonlinearity - signaling,Czachor nonlinearity - signaling,Mielnik nonlinearity - signaling,Simon no-signaling imply linearity} and computational schemes fundamentally much faster than quantum computation \cite{Abrams-Lloyd nonlinearity - fast computation,Aaronson nonlinearity-nonunitary - fast computation}. It is also interesting to investigate the above deep connection between the nonlinearity in the Schr\"odinger equation and the deviation from the Heisenberg uncertainty principle, with the theoretical results that both may imply violations of the second law of thermodynamics \cite{Peres nonlinearity violates 2nd law,Hanggi a deviation from UR violates 2nd}. 

It is remarkable that the linear Schr\"odinger equation of Eq. (\ref{Schroedinger equation}), and the exact form of Heisenberg uncertainty principle of Eq. (\ref{Heisenberg-Kennard UR}), are regained for a specific estimation scheme with the estimation error taking the specific form given by Eq. (\ref{estimation error}) satisfying the principle of estimation independence. On the other hand, other forms of estimation errors violating the principle of estimation independence and/or having unpleasant statistical property from the view of statistical estimation, lead to nonlinear corrections to the Schr\"odinger equation and deviations from the Heisenberg uncertainty principle. Noting that such deviations from linear Schr\"odinger equation and Heisenberg uncertainty principle may be in conflict with the principle of no-signaling and the second law of thermodynamics, or imply implausible computational power, it is natural to ask if the inferential-causality principle of estimation independence together with other reasonable informational constraints, may be upgraded as the axioms to single out uniquely the specific form of estimation error of Eq. (\ref{estimation error}) leading to the standard quantum mechanics. That this might be so is argued in a different work \cite{Agung estimation independence}. Our results also suggest possible deep interlinks between the principle of estimation independence, no-signalling, and the second law of thermodynamics, and other principles used to single out quantum correlation such as information causality \cite{Pawlowski informational approach Tsirelson bound} or data processing inequality \cite{Dahlsten DPI,Wakakuwa GMI,Al-Safi DPI}, worth further investigation in the future.   

The above observation prompts the following question: beside that mentioned in Section \ref{Nonlinear Schroedinger equation and modified Heisenberg uncertainty principle within an estimation scheme}, what kinds of generalizations of the specific estimation scheme of Section \ref{An epistemic reconstruction of quantum mechanics: estimation under epistemic restriction}, comply with the principle of estimation independence, leading to possible nontrivial extensions of the standard quantum mechanics? First, when deriving Eq. (\ref{modified HJM equation}) by imposing the conservation of average energy of Eq. (\ref{conservation of average energy}) leading to the derivation of the Schr\"odinger equation, we have implicitly assumed that the Planck constant $\hbar$, which is the variance of the global random variable $\xi$, is indeed constant in time. One could thus ponder the possibility that $\hbar$ may, though extremely weakly, depend on time, i.e., $\partial_t\hbar\neq 0$. Such an assumption clearly does not violate the principle of estimation independence, and may lead to a weak nontrivial nonlinearity in the Schr\"odinger equation. We may also study the trade-off between the resulting nonlinearity in the Schr\"odinger equation and the possible violations of Born's quadratic law envisioned in Refs. \cite{Valentini nonequilibrium,Aaronson nonlinearity-nonunitary - fast computation}. And, following Valentini's insight in Ref. \cite{Valentini nonequilibrium}, it might be interesting to see the implications of such possible weak temporal fluctuation of $\hbar$ in the early universe. One may also impose additional statistical constraints, reflecting some other symmetries of the statistical estimation problems, when exercising the conservation of average energy of Eq. (\ref{conservation of average energy}). For example, one may assume that some measures of information are (or are not) conserved. Yet another interesting way to generalize quantum mechanics within the operational framework of estimation without violating the principle of estimation independence is to assume that the conservation of trajectories of Eq. (\ref{continuity equation}) is no longer valid as in open systems, or to assume that the conservation of average energy of Eq. (\ref{conservation of average energy}) is somehow violated as in dynamical collapse models \cite{Bassi collapse model review}, which, for example, might be relevant in the cosmological context \cite{Josset violation conservation energy dark energy}. 

Hence, like other operational approaches to reconstruct quantum mechanics \cite{Hardy axioms,D'Ariano generalized probabilitstic theory,Dakic-Brukner axioms,Masanes axioms,Chiribella axioms,Paterek axioms,Chiribella-Spekkens quantum axioms proceedings}, the operational scheme of estimation of momentum given the positions under epistemic restriction discussed in the present work, provides a general epistemic framework encompassing classical, quantum, and a broad class of possible post quantum theories. Note however that, unlike those in Refs. \cite{Hardy axioms,D'Ariano generalized probabilitstic theory,Dakic-Brukner axioms,Masanes axioms,Chiribella axioms,Paterek axioms,Chiribella-Spekkens quantum axioms proceedings}, we have worked directly with the phase space variables so that the transition to classical mechanics is conceptually less painful. Noting this, it is intriguing to investigate possible hybrid interactions between quantum, post-quantum, and classical systems to yet generalize quantum mechanics within the general epistemic framework. For example, a hybrid quantum-classical interaction  \cite{Sudarshan quantum-classical hybrid model,Peres quantum-classical hybrid model,Hall quantum-classical hybrid model} might find applications in developing approximations in computational physics and chemistry \cite{computational quantum chemistry}, for describing nano-mechanical systems in quantum-classical boundary \cite{Aspelmeyer opto-mechanical systems}, and in the study of quantum gravity \cite{Hall witnesing nonclassical gravity,Marletto graviation induced entanglement,Bose gravitation induced entanglement}.     

\begin{acknowledgments}   

This work is partially supported by the Ministry of Education and Culture, and the Ministry of Research and Technology of Republic of Indonesia, under the grant scheme ``Penelitian Dasar Unggulan Perguruan Tinggi (PDUPT),'' and the WCU Program managed by Institut Teknologi Bandung. It is also supported by the John Templeton Foundation (Project No. 43297). The opinions expressed in this publications do not necessarily reflect the views of the John Templeton Foundation. The Authors would like to thank the anonymous Referees for the constructive comments and recommendations, and Daniel Rohrlich, Katsuhiro Nakamura, and Husin Alatas for useful discussions. 

\end{acknowledgments}   

\appendix 

\section{The reconstruction of quantum mechanics within the operational scheme of estimation under epistemic restriction: single and double slits experiments \label{A concrete illustration of the epistemic reconstruction of quantum mechanics within the operational scheme of estimation under epistemic restriction}}

Consider a beam of particles passing through a screen with a single slit, one by one, followed by the detection of the position of the particles (position measurement), e.g., by a second screen at some time $t_0$. Suppose that the agent can control the width of the slit (and possibly some other macroscopic setting parameters such as the average kinetic energy of the particles). Within the model, such a set of macroscopic settings determines a random momentum field $\tilde{p}(q;\xi)$ which, unlike in classical mechanics, irreducibly parameterizes the allowed distribution of position $\rho_{\tilde{p}}(q)$. In this preparation setting, the main idea in the epistemic reconstruction of quantum mechanics based on the estimation scheme is that the agent wants to estimate the underlying momentum field associated with the above macroscopic setting, given information on the conjugate position.   

The estimation of the momentum at a given position is carried out in a ``naive classical'' way as follows \cite{Wiseman Bohmian velocity from naive weak value measurement}. (See also Sec. \ref{An epistemic reconstruction of quantum mechanics: estimation under epistemic restriction} of the main text.) Consider a sub-ensemble of the particles that are detected at the screen to be at $q(t)$ at time $t$, where different trajectories of the particles correspond to different fluctuations of $\xi$. For each of the particle in the sub-ensemble, we make a sufficiently weak measurement of position at $t-\Delta t$ without appreciably altering the subsequent dynamics of the particle, yielding $q(t-\Delta t)$, where $\Delta t$ is extremely small. The velocity along the trajectory at $q(t)$ is then computed in the conventional way, i.e., by taking the difference between $q(t)$ and $q(t-\Delta t)$ and dividing it by $\Delta t$, from which one also obtains the momentum $\tilde{p}(q;\xi)$. Note that, because of the fluctuation of $\xi$, each single repetition of such momentum measurement must give random outcome. To overcome this uncertainty, we then define the estimator for the momentum at $q(t)$ by taking the average of the above measurement outcomes over the sub-ensemble of trajectories passing through $q(t)$, i.e., by averaging over $\xi$ as in Eq. (\ref{estimate of momentum based on mean average}). Based on this estimate, to have a smooth correspondence with the classical relation of Eq. (\ref{HJ condition}), we then construct a real-valued function $S(q)$ satisfying Eq. (\ref{weakly unbiased best estimator}). Moreover, from the distribution of the position $\rho_{\tilde{p}}(q)$ obtained in measurement, the single-shot estimation error is assumed to take the form given by Eq. (\ref{estimation error}). 

As an example, suppose that the agent's estimate of the momentum field along the direction perpendicular to the direction of the beam obtained operationally by following the above scheme, is given by $p_o$ independent of $q$. Then, following Eq. (\ref{weakly unbiased best estimator}), the agent associates a real valued function $S(q)$ satisfying $\partial_qS=p_o$ to give $S(q)=p_oq$. Moreover, suppose the distribution of position of the particles is given by a Gaussian distribution $\rho_{\tilde{p}}(q)=\frac{1}{\sqrt{2\pi\sigma_q^2}}e^{-\frac{(q-q_o)^2}{2\sigma_q^2}}$ with a variance $\sigma_q^2$ assumed to be determined by the width of the slit. Then, from Eq. (\ref{estimation error}), the agent should assign a single-shot estimation error $\epsilon_p(q;\xi)=\frac{\xi}{2}\partial_q\ln\rho_{\tilde{p}}(q)=-\frac{\xi}{2\sigma_q^2}(q-q_o)$ so that the MS estimation error is given by $\mathcal{E}_p^2\doteq\int{\rm d}q{\rm d}\xi(\epsilon_p(q;\xi))^2\chi(\xi)\rho_{\tilde{p}}(q)=\hbar^2/4\sigma_q^2$, where we have used Eq. (\ref{Planck constant}). The above agent's estimation (knowledge) about the momentum field $\tilde{p}(q;\xi)$ at time $t_0$, i.e., the estimator and the estimation error, is then recast compactly into a wave function via $(S(q),\rho_{\tilde{p}}(q))$ defined as in Eq. (\ref{wave function}), i.e., $\psi_0(q)\doteq\sqrt{\rho_{\tilde{p}}(q)}e^{\frac{i}{\hbar}S(q)}=(\frac{1}{2\pi\sigma_q^2})^{1/4}e^{-\frac{(q-q_o)^2}{4\sigma_q^2}+\frac{i}{\hbar}p_oq}$. Such a reconstruction of wave function is in practice similar to the reconstruction of wave function via momentum weak value measurement discussed in Ref. \cite{Agung ERPS distribution}. Hence, by decreasing (increasing) the width of the slit, which means decreasing (increasing) $\sigma_q^2$ implying sharper (poorer) knowledge of the position, then $\mathcal{E}_p^2$ increases (decreases) so that the agent's estimation about the momentum becomes poorer (sharper); and this leads to a narrower (broader) Gaussian wave function. As a limiting case, suppose that the slit is infinitely wide, so that $\sigma_q^2\rightarrow\infty$, implying an infinitely poor knowledge of the position. In this case, the agent's estimate of the momentum $p_o$ is infinitely sharp with a vanishing MS error, i.e.,  $\mathcal{E}_p^2\rightarrow 0$, and the agent should assign a plane wave function $\psi_0(q)\sim e^{ip_oq/\hbar}$ to her preparation.  

Now, suppose that the agent postpones the detection of the position of the particle at some later time $t_1 > t_0$. The question is then, given her estimation about the system | i.e., the estimator for the underlying momentum field and the associated estimation error | at time $t_0$ represented by $\psi_0(q)$, how should she rationally update her estimation at time $t_1$? Suppose further that during the time $t_0\le t\le t_1$, the agent does not make any selection of trajectories so that she does not have new information about her system. The only thing that the agent knows is that the system evolves according to some Hamiltonian. In this case, since she does not make a selection of trajectories, her estimation at time $t_1$ must be updated by respecting the conservation of trajectories and average energy. We have shown in the manuscript that in this case, the wave function representing the agent's estimation (i.e., the estimator and the estimation error) has to be updated following the Schr\"odinger equation, either linear or nonlinear, depending on the assumed exact form of the estimation errors. See Sec. III B for the detailed derivation. The linear Schr\"odinger equation is regained when the estimation error takes the specific from of Eq. (\ref{estimation error}). 

Suppose instead that at some time $t_M$, $t_0<t_M<t_1$, the agent makes a measurement on some physical quantities. Such a measurement in practice corresponds to a selection of a sub-ensemble of trajectories associated with the measurement outcome. Namely, in general, a measurement of a physical quantity with an outcome $o$, corresponds to the selection of a sub-ensemble of trajectories leading to the unambiguous assignment of $o$ (see Ref. \cite{Agung epistemic interpretation}). As a concrete example, consider the paradigmatic which-way measurement by inserting a screen with a double slits, in the middle between the screen with a single slit and the detecting screen. In this case, the outcome ``upper''-way (``lower''-way) corresponds to the selection of those sub-ensemble of trajectories which pass through the upper (lower) slit. Hence, the measurement is carried out by selecting a particular subset of trajectories, so that the conservation of trajectories and average energy no longer apply. Accordingly, the agent's estimation, represented by the wave function, no longer follows the Schr\"odinger equation; instead it must follow a wave function collapse reflecting the Bayesian updating due to the new information associated with the selected sub-ensemble of trajectories \cite{Agung epistemic interpretation}. This is the reason why, in standard quantum mechanics, such a which-way measurement demolishes (suppresses) the interference pattern at the detecting screen. That is, since the linear Schr\"odinger equation is no more valid, the superposition principle no longer applies.    

\section{Proof that the estimator of Eq. (\ref{weakly unbiased best estimator}) with the estimation error of Eq. (\ref{generalized estimation error}) minimizes the mean-squared error \label{proof of best estimator}}

First, given information on $q$, assume a general estimator $T_{p_j}(q)$ for the momentum field $\tilde{p}_j(q;\xi)$, $j=1,\dots,N$, and compute the associated MS estimation error, to obtain, for each degree of freedom $j$: 
\begin{eqnarray}
&&\int{\rm d}q{\rm d}\xi\big(\tilde{p}_j(q;\xi)-T_{p_j}(q)\big)^2\chi(\xi)\rho_{\tilde{p}}(q)\nonumber\\
&=&\int{\rm d}q{\rm d}\xi~\tilde{p}_j(q;\xi)^2\chi(\xi)\rho_{\tilde{p}}(q)+\int{\rm d}q\big(-2T_{p_j}(q)\partial_{q_j}S(q)\nonumber\\
&+&T_{p_j}(q)^2\big)\rho_{\tilde{p}}(q)\nonumber\\
&=&\int{\rm d}q{\rm d}\xi~\tilde{p}_j(q;\xi)^2\chi(\xi)\rho_{\tilde{p}}(q)\nonumber\\
&+&\int{\rm d}q\Big(\big[T_{p_j}(q)-\partial_{q_j}S(q)\big]^2-\partial_{q_j}S(q)^2\Big)\rho_{\tilde{p}}(q), 
\end{eqnarray}
where we have inserted Eq. (\ref{generalized fundamental epistemic decomposition}) and used $\overline{\xi}=0$ in the first equality to obtain the second term on the right hand side. It is then clear that the MS error reaches its minimum when 
\begin{eqnarray}
T_{p_j}(q)=\partial_{q_j}S(q), \nonumber
\end{eqnarray}
$j=1,\dots,N$, as claimed in the main text. In general, one can show that the unbiased estimator for momentum given positions with minimum MS error, is given by the conditional average of momentum given positions, i.e., $T_{p_j}(q)\big|_{\{{\rm min.MS.error}\}}=\overline{p}_j(q)=\int{\rm d}\xi\tilde{p}_j(q;\xi)\chi(\xi)=\partial_{q_j}S(q)$, $j=1,\dots,N$.

\section{The derivation of Eq. (\ref{modified HJM equation 0}) \label{derivation of HJM equation}}

Taking the total derivative of Eq. (\ref{average energy}) with respect to time, one first gets
\begin{eqnarray}
&&\frac{{\rm d}}{{\rm d}t}\braket{H}_{\{S,\rho_{\tilde{p}}\}}\nonumber\\
&=&\int{\rm d}q\Big(\frac{\delta\braket{H}_{\{S,\rho_{\tilde{p}}\}}}{\delta\rho_{\tilde{p}}(q)}\frac{\partial\rho_{\tilde{p}}(q)}{\partial t}+\frac{\delta\braket{H}_{\{S,\rho_{\tilde{p}}\}}}{\delta S(q)}\frac{\partial S(q)}{\partial t}\Big)\nonumber\\
&=&\sum_{j=1}^N\int{\rm d}q\Big(\Big[\frac{(\partial_{q_j}S)^2}{2m_j}-\frac{\hbar^2}{2m_j}\frac{\partial_{q_j}^2\sqrt{\rho_{\tilde{p}}}}{\sqrt{\rho_{\tilde{p}}}}\nonumber\\
&+&V(q)+\frac{\delta  D_f(\rho_{\tilde{p}})}{\delta \rho_{\tilde{p}}}\Big]\partial_t\rho_{\tilde{p}}(q)-\partial_{q_j}\Big(\rho_{\tilde{p}}\frac{\partial_{q_j}S}{m_j}\Big)\partial_tS\Big),
\label{modified HJM equation 0 Appendix}
\end{eqnarray}
where we have used the following result for functional derivatives:
\begin{eqnarray}
&&\frac{\delta}{\delta\rho_{\tilde{p}}(q)}\int{\rm d}q'\frac{1}{8}\Big(\frac{\partial_{q'}\rho_{\tilde{p}}(q')}{\rho_{\tilde{p}}(q')}\Big)^2\rho_{\tilde{p}}(q')\nonumber\\
&=&-\frac{1}{8}\Big(\frac{\partial_q\rho_{\tilde{p}}}{\rho_{\tilde{p}}}\Big)^2-\frac{1}{4}\partial_q\Big(\frac{\partial_q\rho_{\tilde{p}}}{\rho_{\tilde{p}}}\Big)\nonumber\\
&=&\frac{1}{8}\Big(\frac{\partial_q\rho_{\tilde{p}}}{\rho_{\tilde{p}}}\Big)^2-\frac{1}{4}\frac{\partial_q^2\rho_{\tilde{p}}}{\rho_{\tilde{p}}}=-\frac{1}{2}\frac{\partial_q^2\sqrt{\rho_{\tilde{p}}}}{\sqrt{\rho_{\tilde{p}}}},
\label{quantum decomposition Appendix 1}
\end{eqnarray}
and
\begin{eqnarray}
\frac{\delta}{\delta S(q)}\int{\rm d}q'\frac{\big(\partial_qS(q')\big)^2}{2m}\rho_{\tilde{p}}(q')=-\partial_q\Big(\rho_{\tilde{p}}\frac{\partial_qS}{m}\Big). 
\label{quantum decomposition Appendix 2}
\end{eqnarray}
Noting Eq. (\ref{continuity equation}), the last term in the fourth line of Eq. (\ref{modified HJM equation 0 Appendix}) becomes $\partial_t\rho_{\tilde{p}}\partial_tS$, so that one obtains 
\begin{eqnarray}
\frac{{\rm d}}{{\rm d}t}\braket{H}_{\{S,\rho_{\tilde{p}}\}}&=&\sum_{j=1}^N\int{\rm d}q\partial_t\rho_{\tilde{p}}(q)\Big[\partial_tS+\frac{(\partial_{q_j}S)^2}{2m_j}+V(q)\nonumber\\
&-&\frac{\hbar^2}{2m_j}\frac{\partial_{q_j}^2\sqrt{\rho_{\tilde{p}}}}{\sqrt{\rho_{\tilde{p}}}}+\frac{\delta  D_f(\rho_{\tilde{p}})}{\delta \rho_{\tilde{p}}}\Big],
\label{modified HJM equation 1 Appendix}
\end{eqnarray}
as claimed in the main text. 

\section{Nonlinear Schr\"odinger equation from action principle\label{Derivation of the nonlinear Schroedinger equation via action principle}}

Here we sketch the derivation of the nonlinear Schr\"odinger equation using the action principle, following Hall and Reginatto's approach \cite{Hall quantum-classical hybrid model}. Assume that $\rho_{\tilde{p}}(q)$ and $S(q)$ constitute a pair of conjugate variables associated with a Hamiltonian functional $\mathcal{H}[\rho_{\tilde{p}}(q),S(q)]$ (it is called as the Hamiltonian ensemble in Ref. \cite{Hall quantum-classical hybrid model}). Hence, the time evolution of the above pair of the conjugate variables satisfy the following pair of canonical Hamilton's equations:
\begin{eqnarray}
\frac{\partial \rho_{\tilde{p}}(q)}{\partial t}=\frac{\delta\mathcal{H}}{\delta S(q)},~~{\rm and}~~\frac{\partial S(q)}{\partial t}=-\frac{\delta\mathcal{H}}{\delta\rho_{\tilde{p}}(q)}. 
\label{canonical Hamiltonian equation}
\end{eqnarray}
Taking the Hamiltonian functional to be equal to the average energy of Eq. (\ref{average energy}) of the statistical model, i.e., $\mathcal{H}[\rho_{\tilde{p}}(q),S(q)]=\braket{H}_{\{S,\rho_{\tilde{p}}\}}$, the pair of equations in Eq. (\ref{canonical Hamiltonian equation}) give respectively the following coupled differential equations:
\begin{eqnarray}
&&\partial_t\rho_{\tilde{p}}=-\sum_{j=1}^N\partial_{q_j}\Big(\frac{\partial_{q_j}S}{m_j}\rho_{\tilde{p}}\Big),\nonumber\\
&&\partial_tS=-\sum_{j=1}^N\frac{(\partial_{q_j}S)^2}{2m_j}+\frac{\hbar^2}{2m_j}\frac{\partial_{q_j}^2\sqrt{\rho_{\tilde{p}}}}{\sqrt{\rho_{\tilde{p}}}}-V(q)-\mathcal{N}_f(\rho_{\tilde{p}}), 
\label{modified HJM eq. and Contintuity eq.}
\end{eqnarray}
where $\mathcal{N}_f$ is defined as in Eq. (\ref{correction to HJM and S equations}). See Appendix \ref{derivation of HJM equation} for the detailed calculations. 

The above pair of coupled equations are just Eqs. (\ref{continuity equation}) and (\ref{modified HJM equation}) of the main text, which can be recast into the nonlinear Schr\"odinger equation of Eq. (\ref{nonlinear Schroedinger equation general}) via the definition of wave function given by Eq. (\ref{wave function}). Note crucially that in the above derivation, the pair of equations in Eq. (\ref{modified HJM eq. and Contintuity eq.}) are obtained via objective least action principle by choosing the correct Hamiltonian functional given by Eq. (\ref{average energy}). By contrast, within our epistemic reconstruction based on the estimation under epistemic restriction, the pair of equations in Eq. (\ref{modified HJM eq. and Contintuity eq.}) are obtained via epistemic-informational constraint of conservation of average energy and trajectories by choosing the correct estimation error of the form in Eq. (\ref{generalized estimation error}). In this sense, the conservation of average energy and trajectories may provide an epistemic interpretation of the apparently objective principle of least action in terms of estimation of momentum given positions.

\appendix 

\end{document}